\documentclass{jfm}
\usepackage{graphicx}
\usepackage{epstopdf,epsfig}
\usepackage{amsfonts,amssymb,amsmath,mathbbol} 
\usepackage{blindtext}
\usepackage{caption}
\usepackage{subcaption}

\usepackage{xcolor} 
\usepackage{color} 
\usepackage{float}
\usepackage{natbib}
\usepackage{rotating}
\usepackage{overpic}
\usepackage{lineno}
\usepackage{multirow}
\usepackage{array}
\usepackage{comment}
\usepackage{csquotes}
\newcommand{\RomanNumeralCaps}[1]

\def\Re{{\it Re}}
\def\Ta{{\it Ta}}
\def\i{{\rm i}}
\def\n{{\bf n}}
\def\v{{\bf v}}

\def\d{{\rm d}}

\shorttitle{Dynamics of wind-stress-driven shear layer}
\shortauthor{Preethi et al.}

\title{A new instability driven by the combined effect of wind stress and rotation in a sheared liquid layer}

\author{S. Preethi,\aff{1}
Ankush Kamboj,\aff{1}
Ramkarn Patne,\aff{2} \corresp{ramkarn@che.iith.ac.in}
 P. A. L. Narayana,\aff{1} \corresp{ananth@math.iith.ac.in} \and
 Kirti Chandra Sahu\aff{2} \corresp{ksahu@che.iith.ac.in}}

\affiliation{\aff{1} Department of Mathematics, Indian Institute of Technology Hyderabad, Kandi - 502284, Telangana,
India
\aff{2}Department of Chemical Engineering, Indian Institute of Technology Hyderabad, Kandi - 502284, Telangana, India}

\begin{document}

\maketitle

\begin{abstract}
We examine the linear stability of a shear flow driven by wind stress at the free surface and rotation at the lower boundary, mimicking oceanic flows influenced by surface winds and Earth's rotation. The linearised eigenvalue problem is solved using the Chebyshev spectral collocation method and a longwave asymptotic analysis. Our results reveal new longwave instability modes that emerge for non-zero rotational Reynolds numbers. It is observed that the most unstable mode, characterised by the lowest critical parameters, corresponds to longwave spanwise disturbances with vanishing streamwise wavenumber. The asymptotic analysis, which shows excellent agreement with numerical results, analytically confirms the existence of this instability. Thus, the present study demonstrates the hitherto unreported combined influence of wind stress and Earth’s rotation on ocean dynamics. 
\end{abstract}



\section{Introduction} \label{sec:intro}

The Ekman layer plays a vital role in geophysical and oceanographic processes by enhancing vertical mixing and redistributing momentum, heat, and salinity \citep{Ekman1905,pedlosky2013geophysical}. It contributes to large-scale circulations, such as ocean gyres and atmospheric cyclones, by influencing surface wind stress and boundary layer dynamics \citep{gill1982atmosphere}. In the ocean, Ekman flow drives upwelling and downwelling currents via Ekman pumping, facilitating nutrient transport and sustaining biologically productive regions \citep{stewart2008introduction}. Surface flow also significantly impacts ocean-atmosphere interactions and the transport of tracers, including nutrients and pollutants \citep{liu2018effect, harcourt2008large, smith1992observed, d2014turbulence}. Coastal systems are likewise influenced by wind-stress-driven instabilities, which play a key role in regulating sediment resuspension and the dispersion of pollutants \citep{cushman2008introduction}. The Ekman layer, characterised by the balance between viscous forces and the Coriolis effect, remains a fundamental framework in geophysical fluid dynamics \citep{holton1979planetary}. Thus, a fundamental understanding of wind-stress-driven instability is essential for improving climate models, enhancing predictions of ocean-atmosphere coupling \citep{vallis2017atmospheric}, and optimising the performance of offshore wind energy systems \citep{archer2005evaluation}. 

\citet{Ekman1905} investigated the effect of Earth’s rotation on ocean currents by analysing flow in a rotating system, where the pressure gradient, Coriolis force, and friction are in balance. He demonstrated that surface currents are deflected by the Coriolis force, leading to a spiralling velocity profile with depth, commonly known as the Ekman spiral. This foundational study established the basis for the modern understanding of ocean circulation and geophysical fluid dynamics. Subsequently, numerous studies have examined the instability of the Ekman layer through both experimental and theoretical approaches \citep{allen2003hydrodynamic}. Experimental investigations \citep{arons1961experimental, faller1963experimental, tatro1967experiments, jarre1996experimental, aelbrecht1999experimental} revealed complex flow phenomena, including concentric vertical jets, wave-like instabilities beyond a critical Reynolds number, and the extension of boundary layer instabilities into the geostrophic interior. \citet{arons1961experimental} confirmed the role of distributed sources and sinks in generating such circulations, consistent with the theoretical predictions of \citet{stern1960instability}. \citet{faller1963experimental} identified distinct instability modes similar to those in rotating disk boundary layers, while \citet{tatro1967experiments} demonstrated how imposed sinks and sources in a rotating basin affect flow stability. \citet{aelbrecht1999experimental} observed that instabilities and transitions to turbulence in both steady and oscillating Ekman layers are significant in shallow tidal seas, influencing sediment transport and biological processes.

\citet{faller1966numerical} theoretically demonstrated that large, coherent vortices can cause significant deviations from the classical Ekman spiral, in agreement with his earlier experimental findings \citep{faller1963experimental}. A series of studies by \citet{lingwood1995absolute, lingwood1996experimental, lingwood1997absolute} explored the onset of absolute instability in the Ekman layer and related rotating boundary layers, including the Bödewadt and von Kármán flows. \citet{hoffmann1998transitions} investigated secondary and tertiary flow states in Ekman–Couette flow, identifying a primary stationary roll instability under purely viscous conditions.
Subsequent work by \citet{lingwood2011effects} and \citet{alveroglu2016effect} examined convective instabilities within the broader BEK (Bödewadt–Ekman–Kármán) family, accounting for Coriolis effects. \citet{cooper2015effect} showed that surface roughness plays a critical role in flow stability, with isotropic roughness exerting a stabilising influence on the stationary mode. Recently, \citet{thomas2023effect} modelled concentric grooves, radial grooves, and isotropic roughness using slip boundary conditions \citep{miklavvcivc2004flow}. Using direct numerical simulations, \citet{shi2017hydrodynamic} showed that turbulence decays in quasi-Keplerian rotating flows due to the suppression of optimal transient growth. Further contributions include the study by \citet{mukherjee2024convective}, who investigated convective instability in Ekman flow over a stretched rotating disk and found that surface stretching stabilises both cross-flow and viscous modes. \citet{tro2024parameterized} analysed the effect of Ekman pumping in tilted $f$-plane rotating convection, where gravity and rotation are misaligned, and observed shifts in the marginal stability curve at low wavenumbers. These findings were later confirmed by \citet{tro2025asymptotic} through asymptotic analysis of Rayleigh–Bénard convection under Ekman pumping.

Beyond Ekman-type flows, rotation alone is also known to significantly influence flow stability. \citet{johnston1972effects} experimentally demonstrated that Coriolis forces can alter local and global instability characteristics in a fully-developed turbulent channel flow. \citet{bidokhti1992structure} investigated turbulence in rotating shear layers and found that anticyclonic flows may initially destabilise before restabilising, while cyclonic flows remain stable. Similarly, \citet{yanase1993rotating,arobone2012evolution} conducted linear stability analyses of planar wakes and mixing layers under rigid-body rotation, showing that strong anticyclonic and cyclonic rotation generally stabilises three-dimensional disturbances. In the boundary layer flow over a rotating disk, the transition to turbulence is found to be initiated by a secondary absolute instability triggered by a preceding primary instability \citep{pier2003finite,mkhinini2013secondary}. Subsequently, \citet{pier2007primary} showed that controlled enhancement and tuning of primary crossflow vortices could delay the transition to turbulence in boundary layer flow over a rotating disk. \citet{ponty2003onset} explored thermal convection between shearing plates subjected to oblique rotation, focusing on regimes with moderate to strong rotation. \citet{oron1997long} presented a comprehensive review of the long-scale evolution of thin liquid films on rotating disks. Recently, \citet{alfredsson2024flows} also reviewed boundary layer instabilities over rotating disks and cones, concluding that cross-flow mechanisms are the primary source of instability in both geometries.

The studies discussed above primarily address the instability of rotational boundary layers. However, under realistic conditions, the influence of surface wind stress is significant and cannot be neglected. Consequently, we next review the literature that addresses the combined effects of wind-stress-driven instability in rotating systems. Using linear stability analysis, \citet{sheremet1997eigenanalysis} investigated the stability of wind-driven ocean circulation in the subtropical region and identified multiple instability modes. Notably, a wall-trapped mode was the first to become unstable as the Reynolds number increased. \citet{meacham1997instabilities,berloff1998stability} examined the linear stability of wind-driven quasi-geostrophic circulation in a rectangular closed basin, while \citet{berloff1999quasigeostrophic} extended this work to analyse the nonlinear evolution of instabilities in western boundary currents, finding that such currents become unstable at moderate Reynolds numbers. \citet{khaledi2009stabilizing} and more recently, \citet{liu2018effect} studied the combined influence of wind shear stress and planetary rotation on turbulence in a system bounded by a free surface and a rotating lower rigid plate.

\citet{miles1960} performed an asymptotic stability analysis of a thin liquid film with a linear velocity profile, bounded by a rigid wall and a free surface, in the high Reynolds number limit. The analysis revealed that the film becomes unstable when both the Reynolds and Weber numbers are sufficiently large. In a related study, \citet{Feldman1957} demonstrated that discontinuities in viscosity or density can destabilise uniform shear flow. Additionally, \citet{benjamin1957wave} investigated a film with a parabolic velocity profile on an inclined plane and found the flow to be unstable for all Reynolds numbers at sufficiently small wavenumbers. Later, \citet{smith1982instability} extended the formulation of \citet{miles1960} by incorporating an additional term in the normal-stress boundary condition, effectively modelling the system as an equivalent two-layer configuration. Unlike the asymptotic approach of \citet{miles1960}, \citet{smith1982instability} numerically solved the full Orr-Sommerfeld equation governing the linear stability of the system and identified a significantly lower critical Reynolds number for destabilisation. Specifically, while \citet{miles1960} predicted a critical Reynolds number of $Re_c = 203$ for the case of zero surface tension, \citet{smith1982instability} obtained a lower value of $Re_c = 78.6$, and with additional refinements to the boundary conditions, the critical Reynolds number was further reduced to $Re_c = 34.2$. \citet{miesen1995hydrodynamic} investigated the stability of shear-driven thin liquid films using a free-surface approximation, in which wave-induced fluctuations in the gas phase are neglected. They showed that the interface conditions of \citet{miles1960}, rather than those of \citet{smith1982instability}, correctly describe the flow stability. Their analysis also resolved a discrepancy between asymptotic solution of  \citet{miles1960} and the numerical results of \citet{smith1982instability}, demonstrating that interface conditions reported by \citet{miles1960} are more appropriate and consistently address inconsistencies in previous studies. Notably, their computed neutral stability curve aligned well with the high-Reynolds-number limit predicted by \citet{miles1960} but not with the predictions by \citet{smith1982instability}. However, their comparisons with configurations, including gas dynamics, indicated that the free-surface approximation is not always valid. More recently, \citet{hossain2023instability} extended the analysis of \citet{smith1982instability}, conducting a linear stability study of a system involving a floating elastic plate over a flow with a slippery lower boundary. They found that the elasticity of the plate had a stabilising effect on the flow. Additionally, \citet{abid2024free} investigated the linear stability of exponential shear profiles in thin liquid layers over inviscid flows, demonstrating that the least stable wavelength scales with the characteristic shear layer thickness.

As discussed above, incorporating rotation into the classical frameworks of \citet{miles1960} and \citet{smith1982instability} enhances their applicability to geophysical flows. In wind-stress-driven free surface systems, rotational effects, especially Coriolis forces, can significantly impact flow stability, particularly when rotating the lower boundary. While Ekman-type flows have been extensively studied, their stability under combined wind stress and rotation remains underexplored, to the best of the authors' knowledge. The present study addresses that gap by examining the linear stability of a shear flow driven by wind stress over a rotating lower boundary. The system comprises a liquid layer sheared by wind at the free surface and supported by a rotating rigid wall. By performing a three-dimensional stability analysis using Chebyshev spectral collocation, we identify a new longwave instability mode, associated with spanwise disturbances of vanishing streamwise wavenumber, which emerges beyond a critical rotational Reynolds number. A supporting longwave asymptotic analysis confirms the numerical results. These findings advance our understanding of rotational shear flows and have important implications for oceanic and atmospheric sciences.

The remainder of this paper is organised as follows. Section~\S\ref{sec:form} presents the mathematical formulation of the base flow and the linear stability equations. The numerical method, based on the Chebyshev spectral collocation technique used to solve the generalised eigenvalue problem, along with validation of the solver, is described in Section~\S\ref{sec:NumMethod}. Section~\S\ref{sec:dis} presents a detailed analysis of the linear stability results, including the longwave asymptotic analysis. Concluding remarks are given in Section~\S\ref{sec:conc}.

\section{Mathematical formulation} \label{sec:form}

\begin{figure}
\centering
\includegraphics[width=0.75\textwidth]{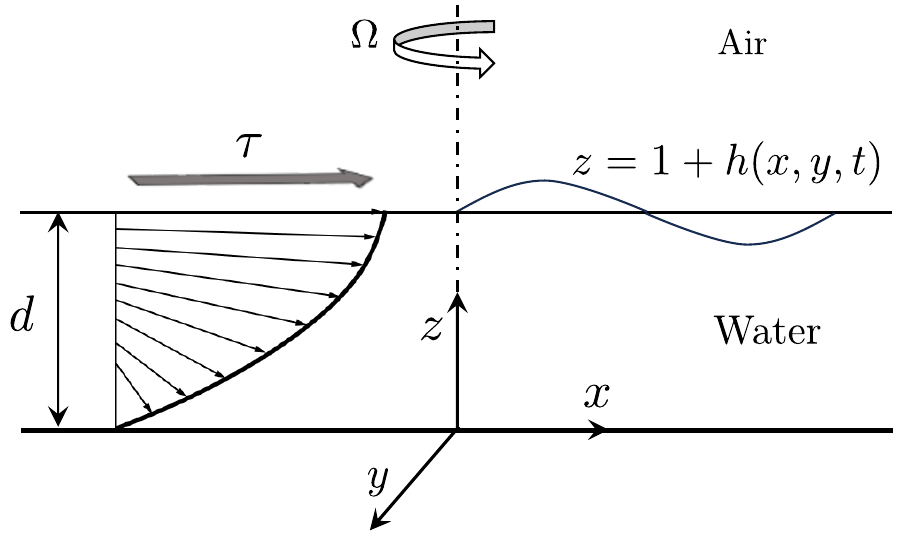}
\caption{Schematic of the flow configuration used to investigate the effect of rotation on a wind-stress-driven shear flow ($\tau$). The lower wall is located at $z = 0$, and the free surface at $z = d$. In dimensionless form, the free surface is expressed as $z = 1 + h(x, y, t)$, where $h(x, y, t)$ denotes the perturbation superimposed on the base state. The lower plane rotates with a constant angular velocity $\Omega$, and the analysis is carried out in a reference frame that rotates with the Earth.}
\label{fig1}
\end{figure}

We examine the linear instability characteristics of a Newtonian, incompressible liquid with density $\rho$, viscosity $\mu$ and height $d$. A constant wind shear stress of magnitude $\tau$ is applied at the free surface, and the lower plane is rotating at a steady angular velocity $\Omega$. The surface tension at the air-water interface is denoted by $\sigma$. Following earlier studies on the linear stability of Ekman flow \citep{lilly1966instability, barcilon1965stability, moberg1984stability}, we consider the flow in a Cartesian coordinate system $(x, y, z)$, schematically illustrated in Figure~\ref{fig1}, where $z = 0$ denotes the lower wall and $z = 1$ the free surface, with the vertical coordinate nondimensionalised by the liquid-layer height. The stability analysis is performed in a reference frame rotating with the Earth. In the present study, the air above the liquid layer is treated as passive, exerting only a constant shear stress at the free surface. To isolate the effect of wind shear on the stability characteristics of the flow, body forces are neglected in the present study.

The non-dimensionalisation of the governing equations is performed using the height $d$ of the liquid layer as the length scale, $\mu/\tau$ as the time scale, $\tau$ as the pressure scale, and $\tau d/\mu$ as the velocity scale. This choice of scaling is motivated by our primary objective to investigate the influence of system rotation and rotational velocity of the lower wall relative to the free surface on the dynamics of the sheared liquid layer. The scaling adopted in this study is the same as that used in \cite{smith1982instability}, which allows us to compare our results with the previous analyses. While the governing equations used in our study are also similar to those in \cite{smith1982instability}, the present investigation incorporates an additional rotational term and extends the analysis to three-dimensional base flow, in contrast to their two-dimensional framework. The resulting dimensionless governing equations are given by
\begin{subequations}
\begin{eqnarray}\label{eq:Dimensionless equations}
\pmb{\nabla} \cdot \boldsymbol{v}=0,
\end{eqnarray}
\begin{eqnarray}\label{eq21b}
\Re \left[\frac{\partial \boldsymbol{v}}{\partial t} + \boldsymbol{v} \cdot \pmb{\nabla}\boldsymbol{v}+2 \Ta \hat{\boldsymbol{k}} \times \boldsymbol{v}\right]= - \pmb{\nabla}{p}+\pmb{\nabla}^2 \boldsymbol{v},
\end{eqnarray}
\end{subequations}
where $\boldsymbol{v} = (u, v, w)$ is the three-dimensional velocity vector, with $u$, $v$, and $w$ representing the velocity components in the streamwise ($x$), spanwise ($y$), and normal ($z$) directions, respectively; $t$ denotes time, $p$ is the pressure, and $\hat{\boldsymbol{k}}$ denotes the unit vector in the $z$ direction. The relevant dimensionless numbers appearing in Eq.~(\ref{eq21b}) are the Reynolds number, $\Re = \rho d^2 \tau / \mu^2$, and the Taylor number, $\Ta = \Omega \mu / \tau$.

We employ the no-slip and no-penetration boundary conditions at the lower wall ($z = 0$), which are given by 
\begin{equation}
u = v = w = 0 \quad \text{at}\;z=0.
\end{equation}
At the free surface ($z = 1+h(x,y,t)$), we impose the continuity of the normal and tangential components of the stress fields, along with the kinematic condition. The normal and tangential stress balance equations at the free surface are given by
\begin{subequations}\label{eq:Boundary Conditions}
\begin{eqnarray}
\sigma_{zz}-2\sigma_{xz}\frac{\partial h}{\partial x}-2\sigma_{yz}\frac{\partial h}{\partial y} = -\frac{S}{Re}\; \pmb{\nabla} \cdot \n,
\end{eqnarray}
\begin{eqnarray}
\sigma_{xz}+h\frac{d\sigma_{xz}}{dz} = 1,
\end{eqnarray}
\begin{eqnarray}
\sigma_{yz}+h\frac{d\sigma_{yz}}{dz} = 0.
\end{eqnarray}
\end{subequations}
Here, $S = \rho d \sigma / \mu^2$ is the dimensionless surface tension, and $\sigma_{ij}$ is the deviatoric stress tensor, which is defined as
\begin{eqnarray}
\sigma_{ij} = -p\delta_{ij} + v_{i,j} + v_{j,i},
\end{eqnarray}
where $\delta_{ij}$ denotes the Kronecker delta, and the Einstein summation convention is applied over the index range $i = 1, 2$ and $3$, representing $x$, $y$ and $z$ directions, respectively. The unit normal and tangential vectors to the free surface are defined as follows:
\begin{eqnarray}
\boldsymbol{n} = \frac{1}{N}(-h_x, -h_y, 1), \quad \boldsymbol{t_1} = \frac{1}{N}(1, 0, h_x), \quad \boldsymbol{t_2} = \frac{1}{N}(0, 1, h_y),
\end{eqnarray}
where $N = (1 + h_x^2 + h_y^2)^{1/2}$. The kinematic boundary condition at the free surface ($z = 1+h(x,y,t)$) is
\begin{eqnarray}\label{eq:KinBC}
w = h_t+u h_x+v h_y.
\end{eqnarray}
Here, the subscripts $t$, $x$, and $y$ denote partial derivatives with respect to time $t$ and the spatial coordinates $x$ and $y$, respectively.
 
\subsection{Base state solution}

We assume the base state, whose linear stability characteristics will be analysed, to be a parallel, fully-developed flow \citep{smith1982instability}. Thus, the base state flow is governed by
\begin{subequations}
\begin{equation} \label{eq27a}
{\d^2 \bar{u} \over \d z^2} + 2\Re_\Omega\bar{v}=0,
\end{equation}
\begin{equation} \label{eq27b}
{\d^2 \bar{v} \over \d z^2}-2\Re_\Omega\bar{u}=0,
\end{equation}
\end{subequations}
where the overbar notation denotes base state quantities, and $\Re_\Omega = \rho d^2 \Omega / \mu$ is the rotational Reynolds number, defined as the product of the Reynolds number ($Re$) and the Taylor number ($\Ta$). The velocity component of the base state in the $z$-direction is zero, i.e., $\bar{w} = 0$. The solution of the base state equations (\ref{eq27a} $-$ \ref{eq27b}) is obtained by applying $\bar{u} = 0$, $\bar{v} = 0$ at $z = 0$, and $d\bar{u}/dz = 1$, $d\bar{v}/dz = 0$ at $z = 1$, under the assumption of fully-developed base flow. The base state velocity field is given by 
\begin{subequations}\label{eq:Basestate V0}
\begin{eqnarray}
\Bar{u}=A\cosh(q z)\sin(q z)+B\cos(q z)\sinh(q z),
\end{eqnarray}
\begin{eqnarray}
\Bar{v}=B\cosh(q z)\sin(q z)-A \cos(q z)\sinh(q z),
\end{eqnarray}
\end{subequations}
where 
\begin{subequations}
\begin{eqnarray}
q=\sqrt{\Re_{\Omega}},
\end{eqnarray}
\begin{eqnarray}
A = \frac{\cos(q)\cosh(q)+\sin(q)\sinh(q)}{q\left[\cos(2q)+\cosh(2q)\right]},
\end{eqnarray}
\begin{eqnarray}
B =  \frac{\cos(q)\cosh(q)-\sin(q)\sinh(q)}{q\left[\cos(2q)+\cosh(2q)\right]}.
\end{eqnarray}
\end{subequations}
By setting $\Ta = 0$ in our formulation, we recover the problem analysed by \cite{smith1982instability}. 

\subsection{Linear Stability Analysis}

We investigate the linear instability characteristics of a sheared liquid layer driven by the combined effects of wind stress and rotation, considering three-dimensional infinitesimal perturbations. The corresponding base state has been discussed in the previous section. To derive the governing linear stability equations, each flow variable is decomposed into a base state and a perturbation (denoted by a prime). For instance, the velocity component in the $x$-direction is expressed as:
\begin{eqnarray}\label{eq:perturbation}
u = \bar{u}(z) + u'(x, y, z, t).
\end{eqnarray}
In the normal mode form, the velocity perturbation in the $x$ and $y$ directions can be expressed as
\begin{eqnarray}
\label{eq:Normalmodes}
u'(x, y, z, t) = \hat{u}(z)\exp\left[\i (kx + my -\omega t)\right],
\end{eqnarray}
where $\i \equiv \sqrt{-1}$, $k$ and $m$ are the streamwise and spanwise wavenumbers, respectively, and $\omega ~(=\omega_r+\i \omega_i)$ is the complex frequency, wherein $\omega_r$ and $\omega_i$ are the real and imaginary parts of $\omega$, respectively. While Eqs.~\eqref{eq:perturbation} and \eqref{eq:Normalmodes} illustrate the decomposition and normal mode form for the velocity component in the $x$ direction, the remaining flow variables, namely $v$, $w$, and $p$, are expressed analogously. Similarly, the perturbation associated with the free surface can be expressed as 
\begin{align}
h(x,y,t)= \hat{h}~\text{exp}\left[\i (kx + my -\omega t)\right].
\end{align}
The temporal growth or decay of a given mode is determined by the sign of $\omega_i$. The flow is linearly unstable if at least one eigenvalue in the eigenspectrum exhibits ${\omega_i} > 0$. Substituting the perturbations into the governing equations, subtracting the base state solution and subsequently imposing the aforementioned normal modes for each perturbed variable in the linearised perturbed system yields:
\begin{subequations}\label{eq:EigenValue}
\begin{eqnarray}
\i k \hat{u}+\i m \hat{v}+D\hat{w}=0,
\end{eqnarray}
\begin{eqnarray}
\left[D^2-(k^2+m^2) -\i Re(k \bar{u}+m \bar{v})\right] \hat{u}+2 Re_{\Omega}\; \hat{v}-Re\;D\bar{u}\;\hat{w}-\i k \hat{p}=-\i Re\;\omega \; \hat{u},~~~~~~
\end{eqnarray}
\begin{eqnarray}
-2 Re_{\Omega} \; \hat{u}+\left[D^2-(k^2+m^2)-\i Re(k \bar{u}+m \bar{v})\right] \hat{v}-Re\;D\bar{v}\;\hat{w}-\i m \hat{p}=-\i Re\;\omega \; \hat{v}, ~~~~~~
\end{eqnarray}
\begin{eqnarray}
\left[D^2-(k^2+m^2)-\i Re(k \bar{u}+m \bar{v})\right] \hat{w}-D\hat{p}=-\i Re\;\omega \; \hat{w}.
\end{eqnarray}
\end{subequations}
Here, $D = \d/\d z$. These equations are then subjected to the following boundary conditions.

At the lower wall ($z=0$):
\begin{subequations}\label{eq:EigenValueBCs}
\begin{eqnarray}
\hat{u}=\hat{v}=\hat{w}=0.
\end{eqnarray}
The boundary conditions for the free surface (at $z=1$) are given below. The kinematic boundary condition is
\begin{eqnarray}
\i(k\Bar{u}+m\Bar{v}-\omega)\hat{h}=\hat{w}. \label{eq:kinmbc}
\end{eqnarray}
The normal and tangential components of the stress balance equations, after applying the kinematic boundary condition \eqref{eq:kinmbc} at $z=1$, yield
\begin{eqnarray}
2 \left [ \left (k \;D\bar{u}+m\;D\bar{v} -(k\bar{u}+m\bar{v})D \right )+\i\frac{S}{Re}\left (k^2+m^2 \right ) \right ]\hat{w} \nonumber \\ + (k\bar{u}+m\bar{v})\hat{p}= \left (-2\;D \hat{w}+\;\hat{p} \right)\omega,
\end{eqnarray}
\begin{eqnarray}
(k\bar{u}+m\bar{v})D \hat{u}+ \left [2\i Re_{\Omega}\bar{v}+\i k(k\bar{u}+m\bar{v}) \right ] \hat{w}=(D \hat{u}+\i k\;\hat{w})\omega ,
\end{eqnarray}
\begin{eqnarray}
(k\bar{u}+m\bar{v})D \hat{v}- \left [2\i Re_{\Omega}\bar{u}-\i m(k\bar{u}+m\bar{v}) \right] \hat{w}=(D \hat{v}+\i m\; \hat{w})\omega.
\end{eqnarray}
\end{subequations}
Eq.~\eqref{eq:EigenValue}, together with the boundary conditions~\eqref{eq:EigenValueBCs}, constitute a generalized eigenvalue problem with eigenvalue $\omega$ and eigenvectors $\mathcal{U}=[\hat{u},\hat{v},\hat{w},\hat{p}]^T$. The details of the numerical procedure are discussed below.

\section{Numerical method and validation}\label{sec:NumMethod}

The generalised eigenvalue problem is solved using the Chebyshev spectral collocation method \citep{canuto1988spectral, sahu2020linear, ankush2023mixed}. In this approach, each dynamical variable is represented as a finite sum of Chebyshev polynomials. The computational domain along the $z$-axis is discretised using collocation points defined by
\begin{equation}
z_j=\frac{1}{2}\left[\cos\left(\frac{\pi j}{N}\right) +1 \right], \quad j=0,1,2,...,N.
\end{equation}
Here, $N$ denotes the total number of collocation points. By discretising the domain along the $z$-axis, the linear stability equations are transformed into a generalised matrix eigenvalue problem, which is given by
\begin{equation} \label{eq:eig}
\mathcal{A\;U} =\omega\;\mathcal{B\;U}.
\end{equation}

\begin{table}
\centering
 \begin{tabular}{c c c} 
 $N$ & $\omega_r$ & $\omega_i$\\ [0.7ex] 
 $10$ & $-0.0061390$ & $5.074 \times 10^{-7}$\\ 
 $20$ & $-0.0061390$ & $5.075 \times 10^{-7}$ \\
 $30$ & $-0.0061390$ & $5.076 \times 10^{-7}$
 \end{tabular}
 \caption{Effect of increasing the order of Chebyshev polynomial $(N)$ on the least stable eigenvalue for $\Re = 0.1$, $\Re_{\Omega} = 0.5$, $k = 0$, $m=0.01$ and $S = 0.01$.}
\label{table:2}
\end{table}

\begin{figure}
\centering
\includegraphics[width=0.5\textwidth]{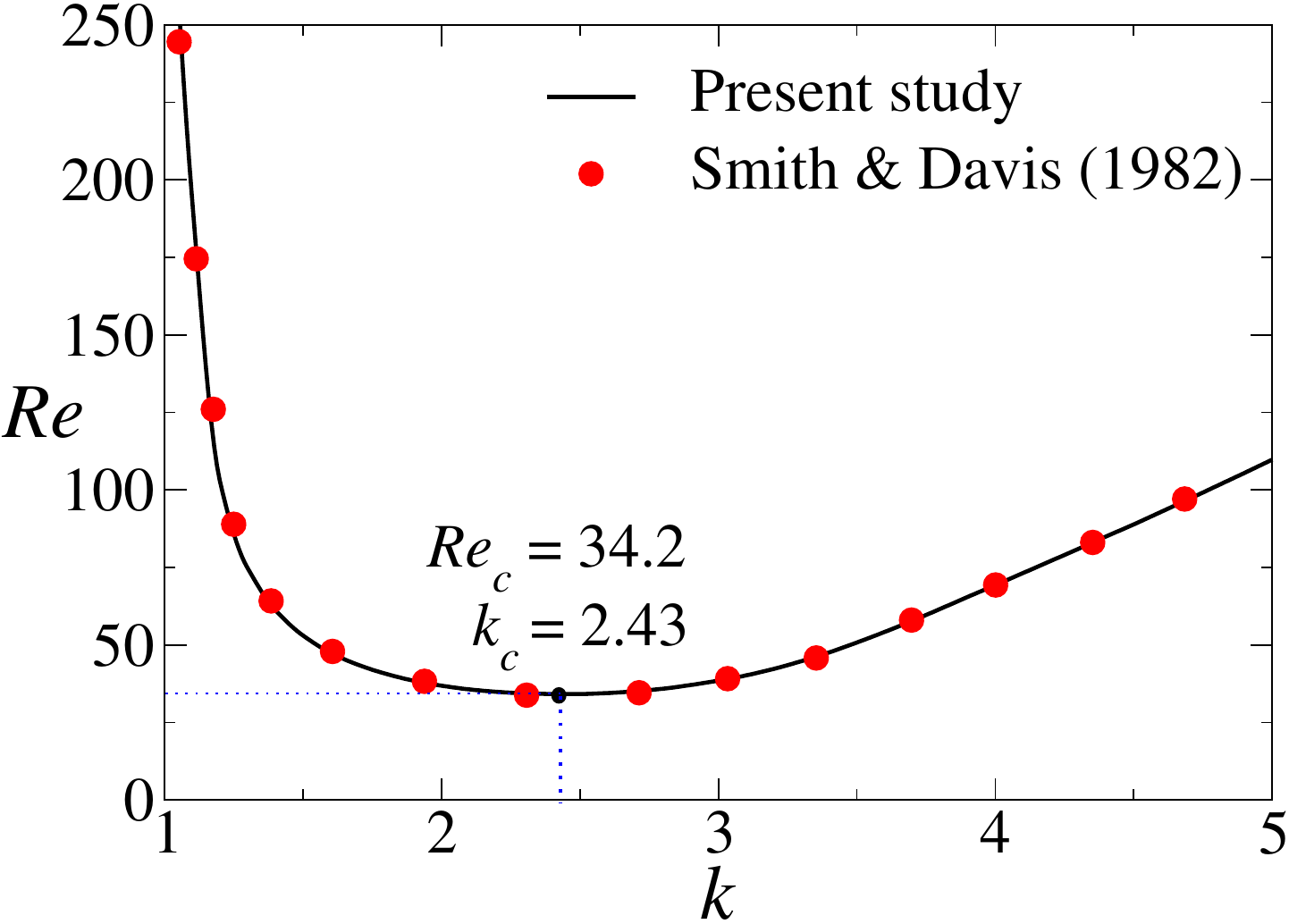}
\caption{Comparison of the neutral stability curve obtained from the present solver with that of \citet{smith1982instability}. The remaining parameters are set as $\Re_{\Omega} = 0$, $m=0$ and $S = 0$. The critical Reynolds number and the critical wavenumber are found to be $\Re_c = 34.2$ and $k_c = 2.43$, respectively, which are in excellent agreement with the values reported by \citet{smith1982instability}.}
\label{fig2}
\end{figure}

The eigenvalue $\omega$ in Eq.~\eqref{eq:eig} is computed using the \textbf{eig} command in \textsc{MATLAB}$^{\circledR}$. We performed a grid convergence test to determine an optimal value of $N$. Table~\ref{table:2} presents the most-unstable eigenvalue obtained using different values of $N$ for a typical set of parameters used in the present study ($\Re = 0.1$, $\Re_{\Omega} = 0.5$, $k = 0$, $m=0.01$ and $S = 0.01$). It can be seen that the most unstable eigenvalues obtained for different values of $N$ are essentially indistinguishable. In view of this, we select $N = 30$ to generate the results presented in the subsequent section. Additionally, the numerical method employed in the present study has been validated by comparing the results obtained from our solver with the previously published findings of \cite{smith1982instability}, as shown in Figure~\ref{fig2}. In order to achieve the results predicted by \cite{smith1982instability}, we set $\Re_{\Omega} = 0$, $m=0$ and $S = 0$, in our formulation. The validation is performed by comparing the neutral stability curve generated by our numerical method with that reported by \cite{smith1982instability}. As seen in figure~\ref{fig2}, the neutral stability curve obtained using our solver shows excellent agreement with that reported by \cite{smith1982instability}. Our analysis predicts the critical Reynolds number $\Re_c = 34.2$ and the critical wavenumber $k_c = 2.43$, which are matching with the values reported by \cite{smith1982instability}, thereby further validating the numerical methodology employed in this study.

\section{Results and discussion} \label{sec:dis}

\subsection{General linear stability analysis} \label{sec:LSA}

We begin the presentation of our results by briefly reviewing the origin and significance of the problem addressed in the present study. \cite{smith1982instability} conducted a linear stability analysis of a thin liquid layer subjected to wind shear stress at the free surface. Their work provided critical insights into the linear stability characteristics of such shear-driven flows in the absence of rotational effects. Extending the earlier work of \cite{miles1960}, \cite{smith1982instability} corrected a missing term in the normal stress boundary condition and numerically solved the associated linear stability equations to determine the critical conditions for instability. Their analysis revealed a significantly lower critical Reynolds number for the onset of instability compared to \cite{miles1960}. Specifically, for zero surface tension ($S = 0$), \cite{smith1982instability} reported a critical Reynolds number of $Re_c = 34.2$, whereas \cite{miles1960} predicted $Re_c = 203$. It is important to note that both studies are applicable primarily to liquid layers, where the influence of Earth’s rotation may be negligible. The present study aims to address this limitation by incorporating the effect of rotation on the dynamics of the sheared liquid layer. In our formulation, the lower boundary rotates with a constant angular velocity $\Omega$, resulting in a modified Ekman-type base flow that is relevant to various geophysical contexts, such as oceanic flows. Within this liquid layer, the velocity direction gradually changes with depth, forming the characteristic Ekman spiral, as illustrated in figure~\ref{fig1}.

\begin{figure}
\centering
\hspace{0.6cm} {\large (a)} \hspace{6.3cm} {\large (b)} \\
\includegraphics[width=0.47\textwidth]{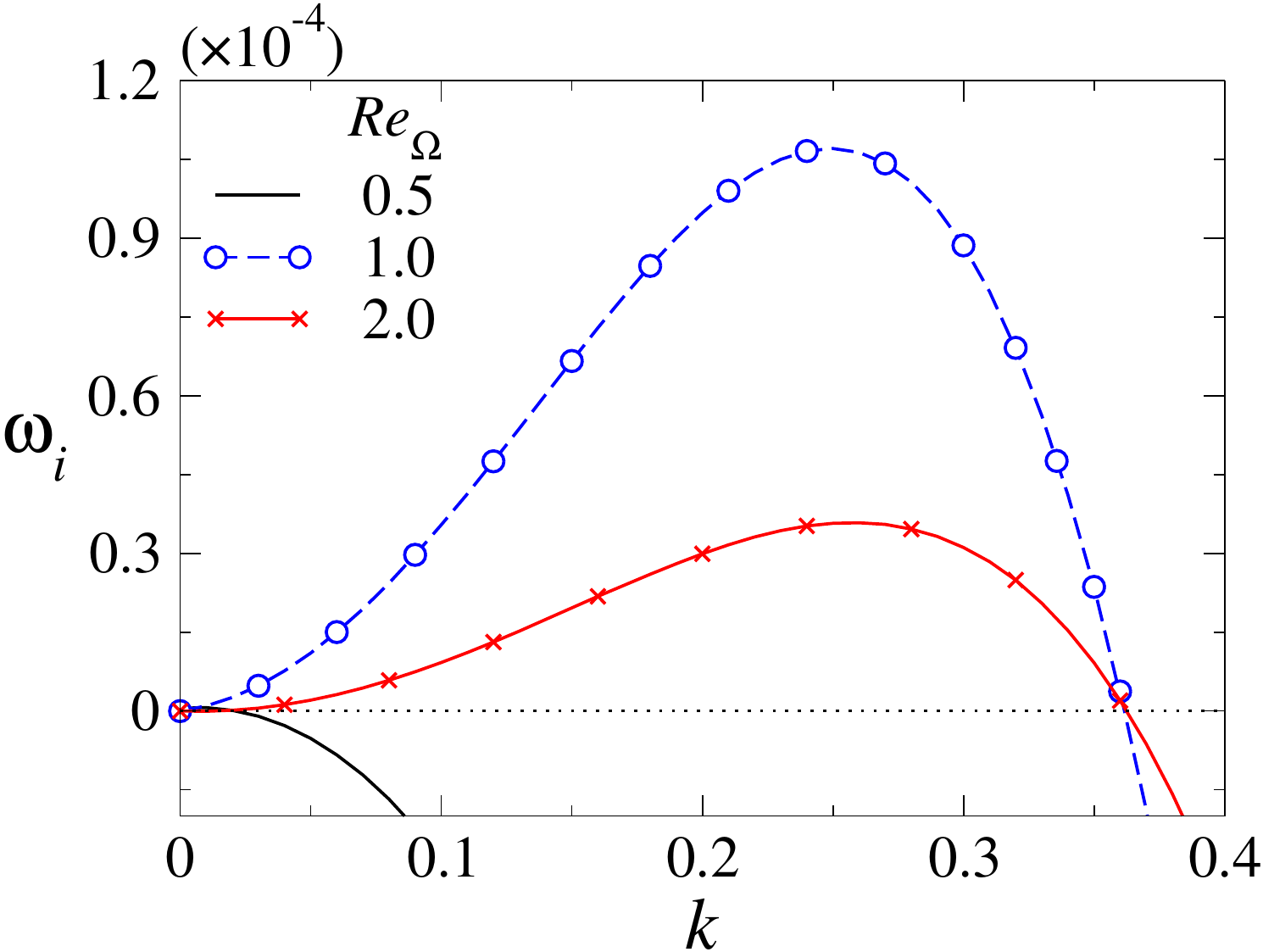} \hspace{5mm} 
\includegraphics[width=0.47\textwidth]{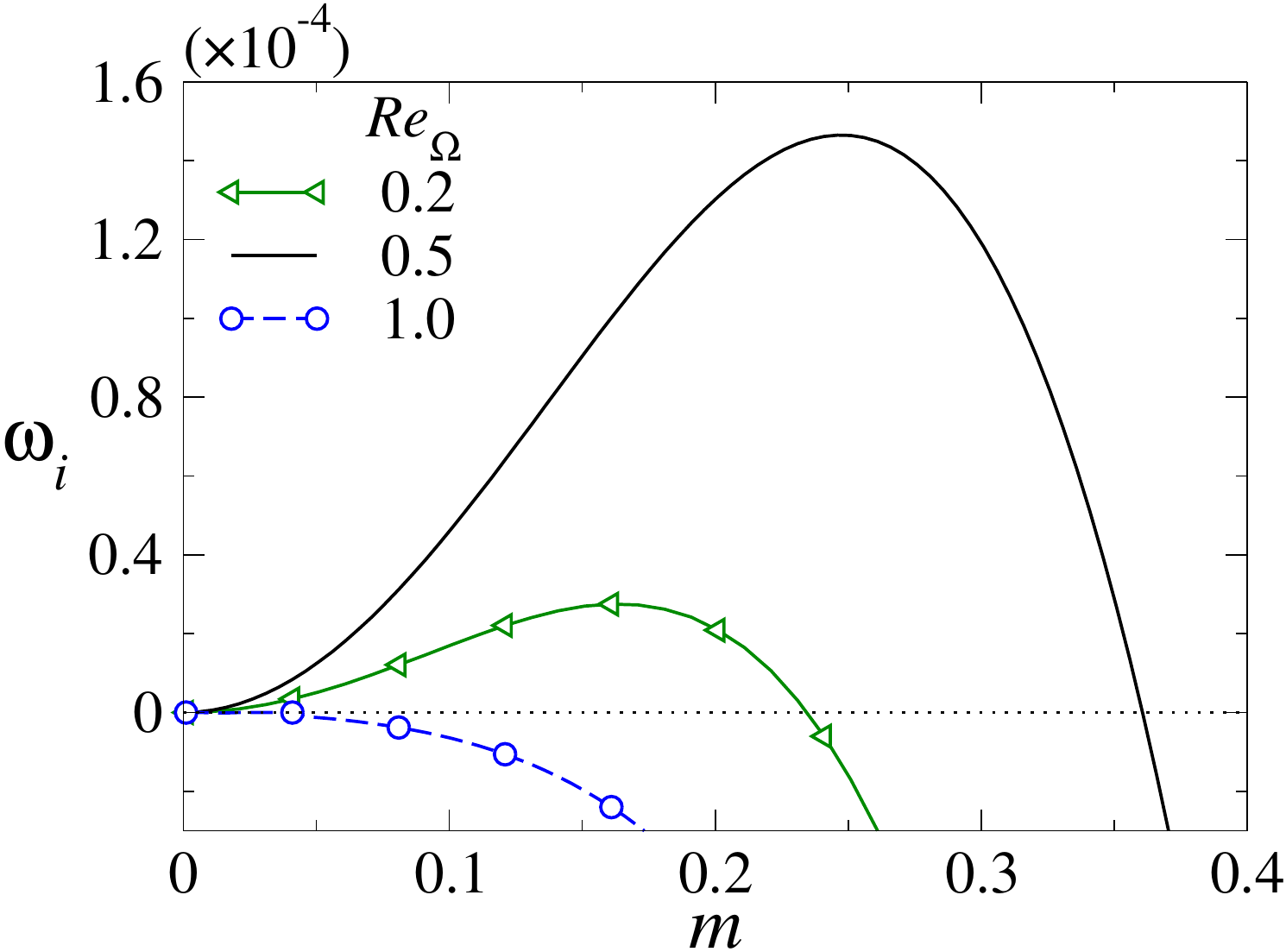}
\caption{(a) The variation of the temporal growth rate of instability $\omega_i$ with $k$ for $m=0.01$, and (b) the variation of $\omega_i$ with $m$ for $k=0$, at different values of $Re_\Omega$. The rest of the parameters are $Re=0.1$ and $S=0.01$.}
\label{fig:DCvaryROm}
\end{figure}

We perform a parametric study by varying the streamwise wavenumber ($k$), spanwise wavenumber ($m$), Reynolds number ($Re$), and rotational Reynolds number ($Re_\Omega$) to examine their influence on the linear stability characteristics of the flow configuration considered in this study. Figures~\ref{fig:DCvaryROm}(a) and \ref{fig:DCvaryROm}(b) show the dispersion curves of the growth rate of the disturbance $\omega_i$ as a function of the streamwise wavenumber $k$ (for $m = 0.01$) and the spanwise wavenumber $m$ (for $k = 0$), respectively, at different rotational Reynolds numbers $Re_\Omega$. The dispersion curves exhibit a paraboloidal shape, with the disturbance growth rate $\omega_i > 0$ over a finite range of streamwise and spanwise wavenumbers, indicating the presence of linear instability. They further reveal a distinct most-dangerous mode, corresponding to the maximum value of $\omega_i$, and cut-off modes, where $\omega_i$ becomes negative. A non-monotonic dependence on $Re_\Omega$ is observed. Specifically, as $Re_\Omega$ increases, the maximum growth rate $\omega_i$ initially amplifies but subsequently decreases at larger values, signifying suppression of the instability at sufficiently high $Re_\Omega$. Notably, the occurrence of $\omega_i > 0$ at small wavenumbers demonstrates the existence of longwave instabilities in the streamwise ($k=0$), spanwise ($m=0$), and oblique ($k \neq 0, m \neq 0$) directions.

\begin{figure}
\centering
\hspace{0.6cm} {\large (a)} \hspace{6.3cm} {\large (b)} \\
\includegraphics[width=0.48\textwidth]{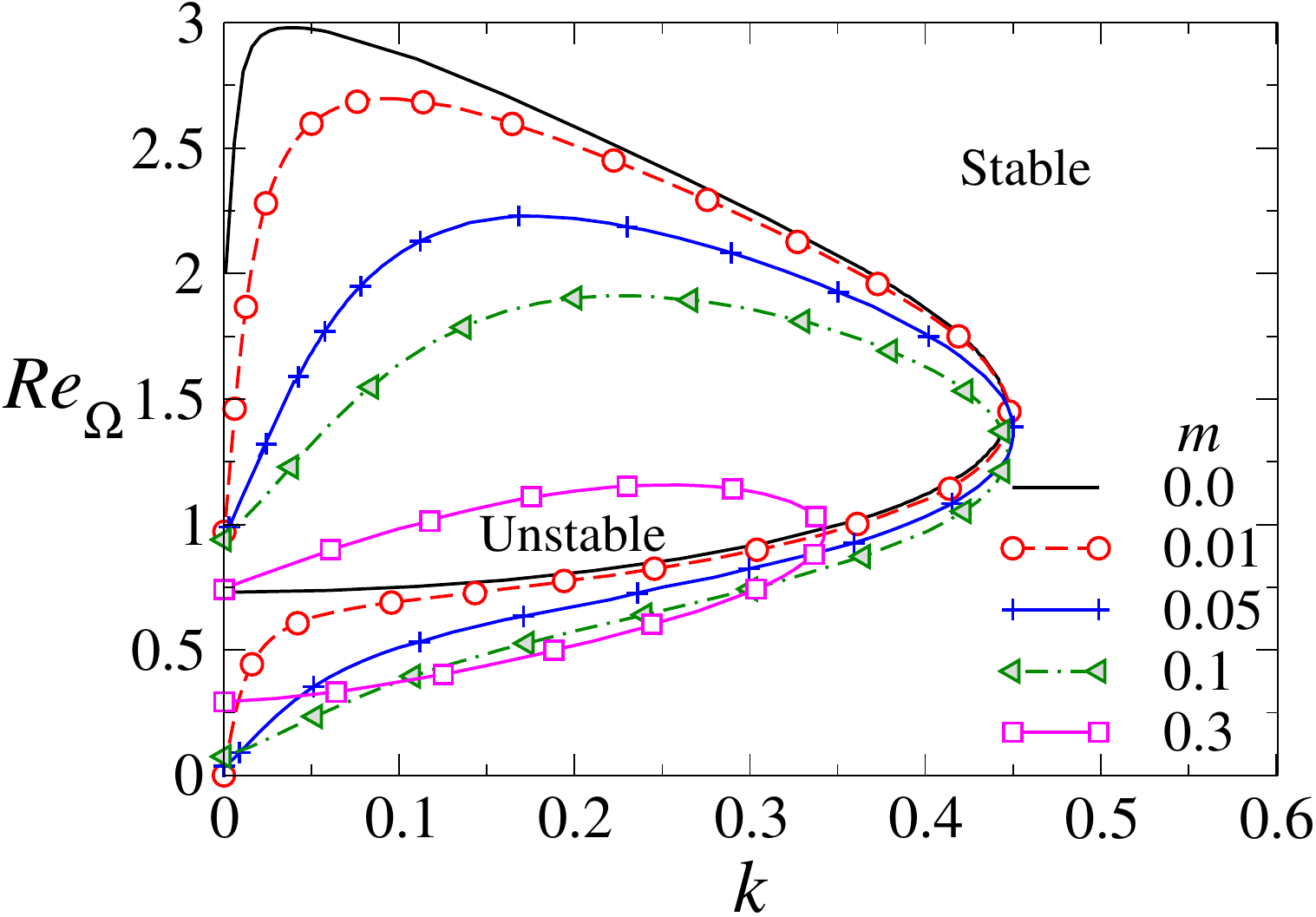} \hspace{2mm} 
\includegraphics[width=0.475\textwidth]{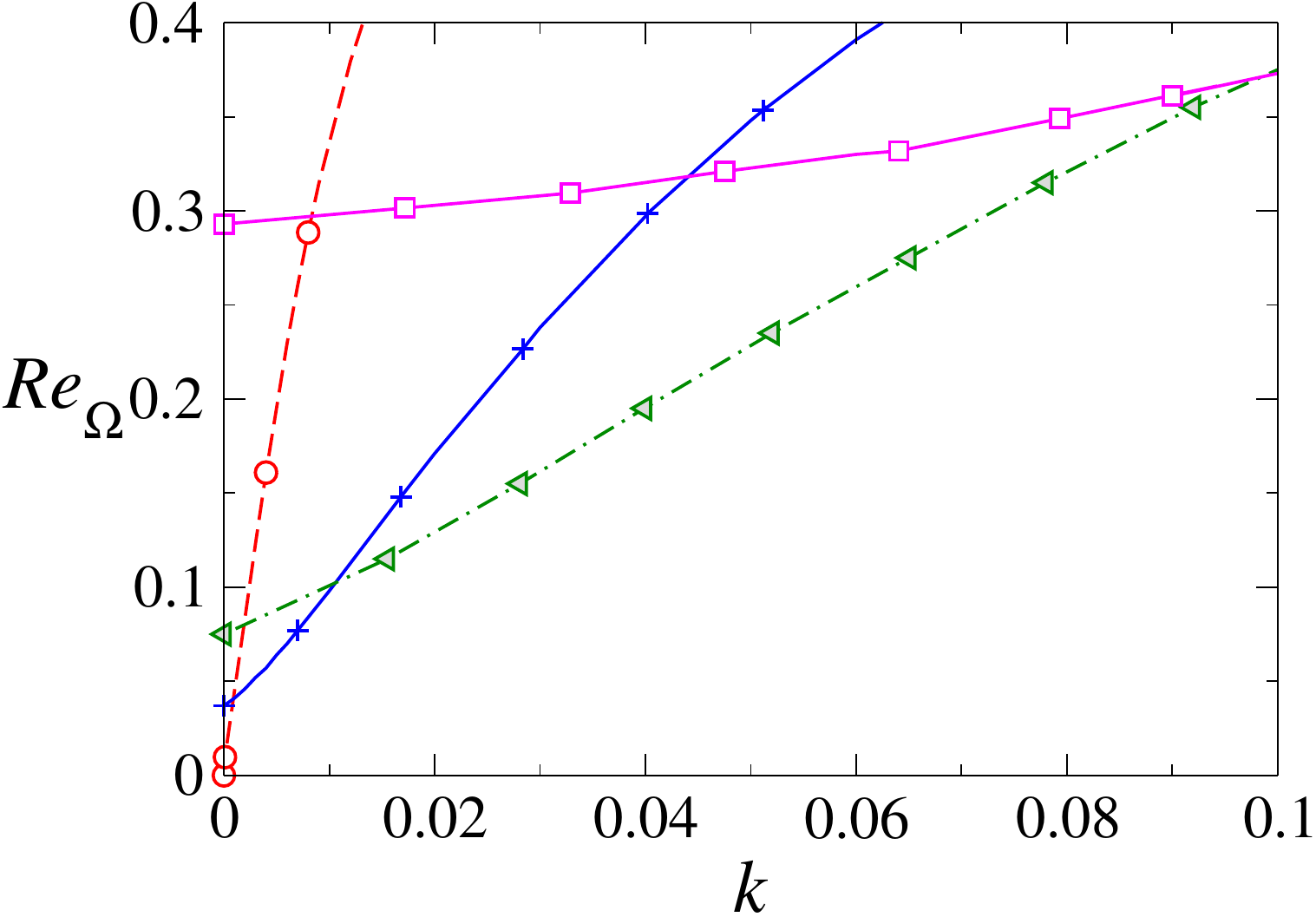}
\caption{Neutral stability curves in the $Re_\Omega - k$ plane for different values of $m$, delineating the stable and unstable regions. Panel (b) presents a magnified view of the bottom-left corner of panel (a). The rest of the parameters are $Re=0.1$ and $S=0.01$.}
\label{fig:NSVarymkvsReOmV0}
\end{figure}

It is even more interesting to examine the neutral stability curves for various parameter combinations to identify the dominant mode, defined as the one with the lowest critical parameter value. Figure~\ref{fig:NSVarymkvsReOmV0}(a) presents the neutral curves in the $(Re_\Omega, k)$ plane for different spanwise wavenumbers $m$, while Figure~\ref{fig:NSVarymkvsReOmV0}(b) shows an enlarged view of the bottom-left region of Figure~\ref{fig:NSVarymkvsReOmV0}(a), allowing us to determine whether the most unstable mode is streamwise, spanwise, or oblique. The results demonstrate that perturbations with $m \neq 0$ destabilise at lower $Re_\Omega$ compared to those with $m = 0$. In particular, Figure~\ref{fig:NSVarymkvsReOmV0}(b) reveals that for $m = 0.01$ and $k = 0$, the critical $Re_\Omega$ approaches zero, implying instability for nonzero rotation. This behaviour resembles the instability in viscosity-stratified flows reported by \cite{yih67}, where the flow becomes unstable at arbitrarily small Reynolds numbers. Moreover, for $m > 0.05$, the critical $Re_\Omega$ increases, indicating that modes with $0 < m < 0.1$ and $k = 0$ correspond to the dominant instability. The predicted instability vanishes for $m > 0.3$, resulting in the absence of neutral stability curves in this range. In summary, the most unstable mode arises for $\Re_\Omega \neq 0$, is characterized by a small but finite spanwise wavenumber and vanishing streamwise wavenumber, and therefore corresponds to a spanwise longwave mode.

\begin{figure}
\centering
\hspace{0.6cm} {\large (a)} \hspace{6.3cm} {\large (b)} \\
\includegraphics[width=0.49\textwidth]{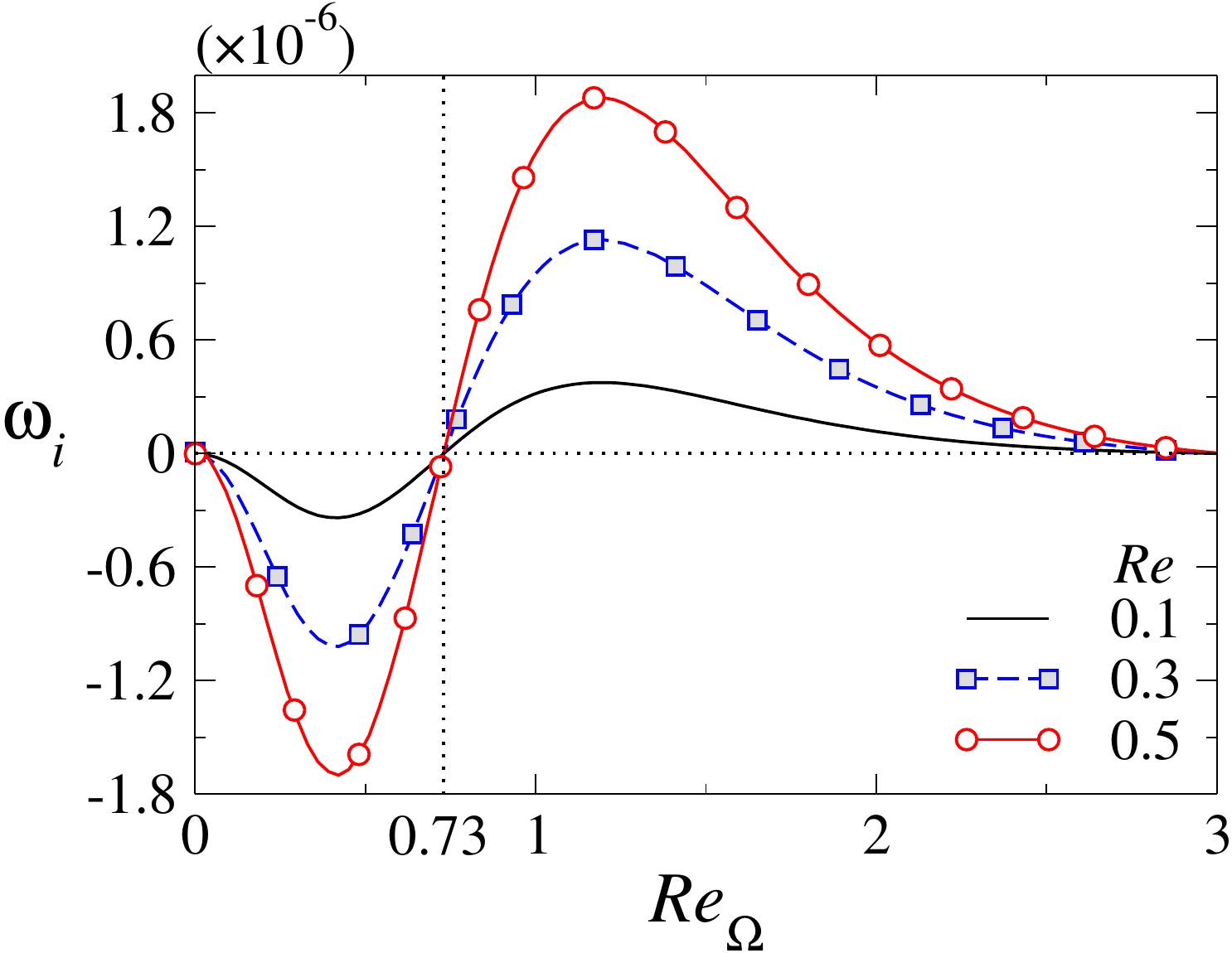} \hspace{2mm} 
\includegraphics[width=0.48\textwidth]{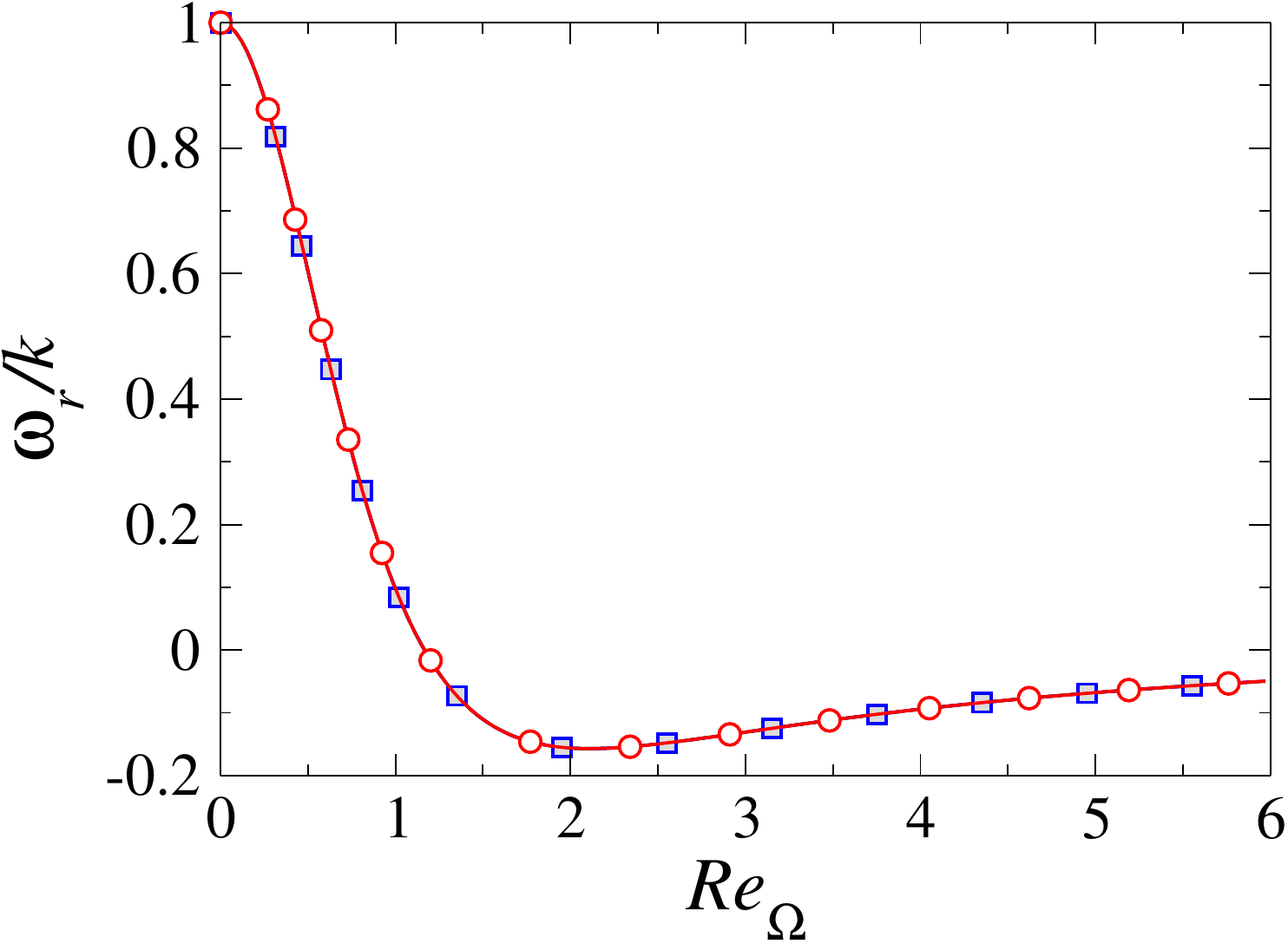}
\caption{Variation of (a) $\omega_i$ and (b) $\omega_r$ with $Re_\Omega$ for different values of $Re$. Here, $S = 0.01$. We set $k = 0.01$ and $m = 0$ to examine the streamwise longwave instability.}
\label{fig:DCReOmvswiwrk001}
\end{figure}

\begin{figure}
\centering
\hspace{0.6cm} {\large (a)} \hspace{6.3cm} {\large (b)} \\
\includegraphics[width=0.48\textwidth]{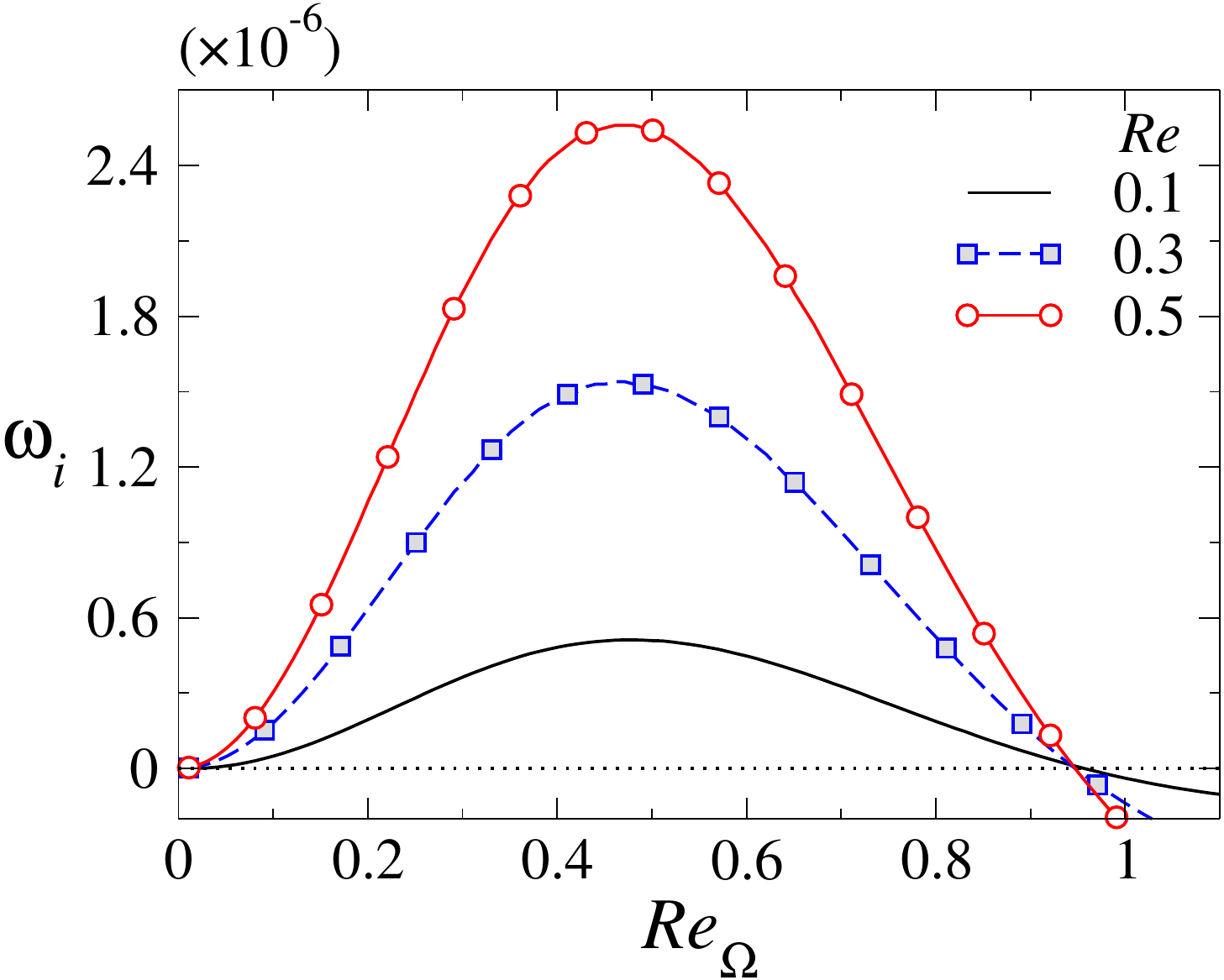} \hspace{2mm} 
\includegraphics[width=0.48\textwidth]{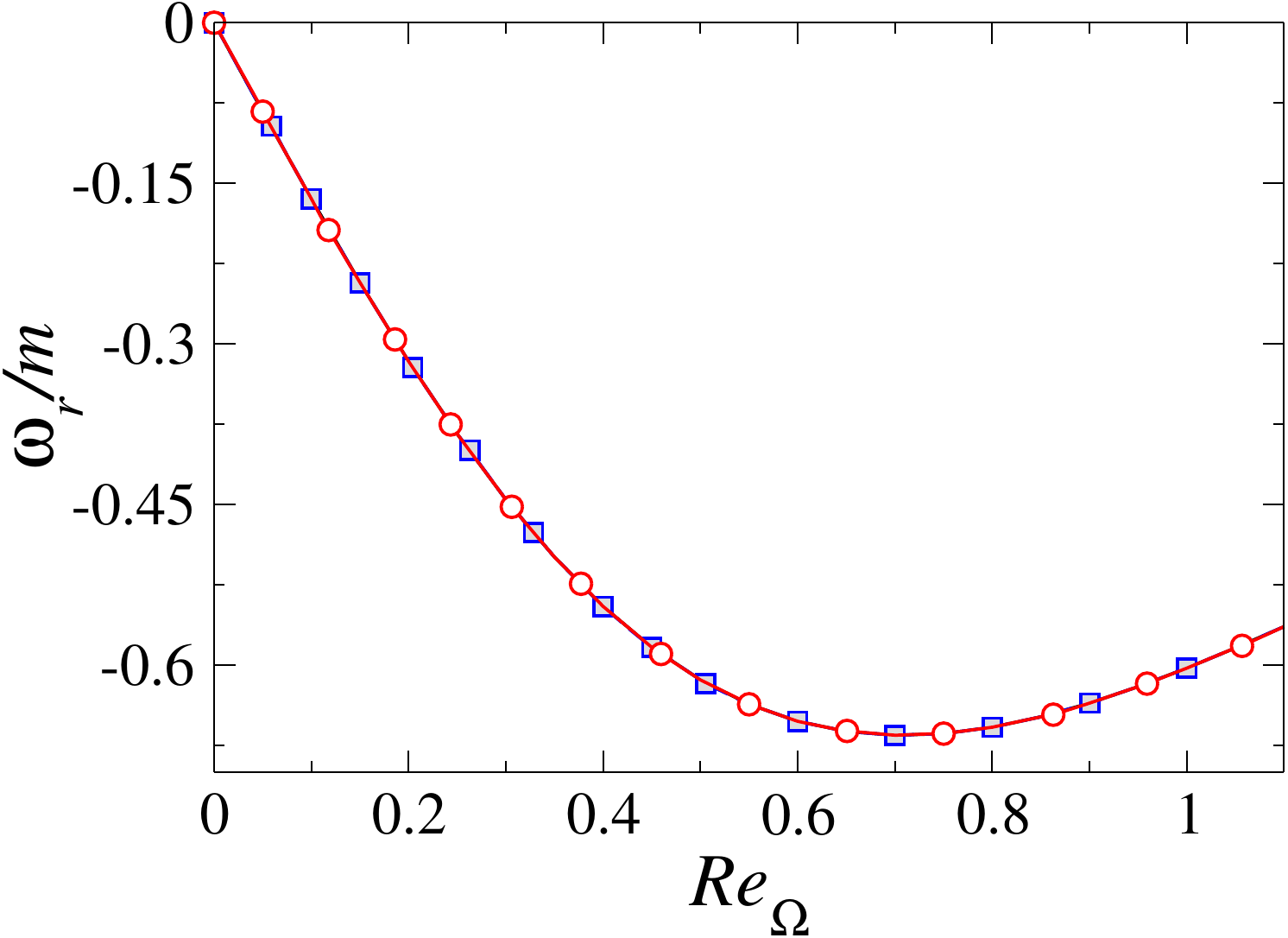}
\caption{Variation of (a) $\omega_i$ and (b) $\omega_r$ with $Re_\Omega$ for different values of $Re$. Here, $S = 0.01$. We set $k = 0$ and $m = 0.01$ to analyse longwave instability.}
\label{fig:DCvaryReROmvswiwr}
\end{figure}

To gain further insight into the longwave instability, Figures~\ref{fig:DCReOmvswiwrk001}(a,b) depict the variation of $\omega_i$ and $\omega_r/k$ for the most unstable disturbance with the rotational Reynolds number ($Re_\Omega$) at different values of $Re$, for the streamwise longwave mode ($k = 0.01$, $m = 0$). It can be seen that the onset of instability, indicated by $\omega_i$ crossing zero, occurs at $Re_\Omega \approx 0.73$ for all values of $Re$, indicating that the flow is stable for $Re_\Omega \leq 0.73$ and unstable beyond this threshold. Figures~\ref{fig:DCvaryReROmvswiwr}(a,b) present the variations of $\omega_i$ and $\omega_r/m$ for the most unstable disturbance with $Re_\Omega$ in the spanwise longwave mode ($k = 0$, $m = 0.01$), where instability arises even at very small $Re_\Omega$. In both cases, the critical $Re_\Omega$ corresponding to the maximum $\omega_i$ is independent of $Re$. Although not shown, the flow remains neutrally stable ($\omega_i=0$) for $Re=0$, indicating that inertia is necessary for the onset of the longwave instability observed in the present study. Notably, rotation leads to instability even at small wavenumbers, a behaviour not captured by the earlier study of \cite{smith1982instability}.  

This raises an important question: do the longwave streamwise and spanwise instabilities arise from newly introduced unstable eigenvalues, or are they existing modes destabilised by rotation? Addressing this question is crucial for understanding the origin of the new instabilities. To investigate, consider first the streamwise mode for which Figure~\ref{fig:DCReOmvswiwrk001}(b) shows the variation of $\omega_r/k$ with $Re_\Omega$. It can be seen in Figure~\ref{fig:DCReOmvswiwrk001}(b) that at $Re_\Omega = 0$, the phase speed of the perturbations ($\omega_r/k$) is unity, corresponding to the neutrally stable free-surface mode. As $Re_\Omega$ increases, $\omega_r/k$ gradually decreases, reaching approximately $\omega_r/k \approx 0.3364$ at the critical $Re_\Omega = 0.73$. This critical value coincides with the onset of instability shown in Figure~\ref{fig:DCReOmvswiwrk001}(a), where $\omega_i$ becomes positive for different $Re$. These observations indicate that the free-surface mode, driven by wind stress, is destabilised by rotation, giving rise to the newly observed longwave streamwise instability.

\begin{figure}
\centering
\includegraphics[width=0.5\textwidth]{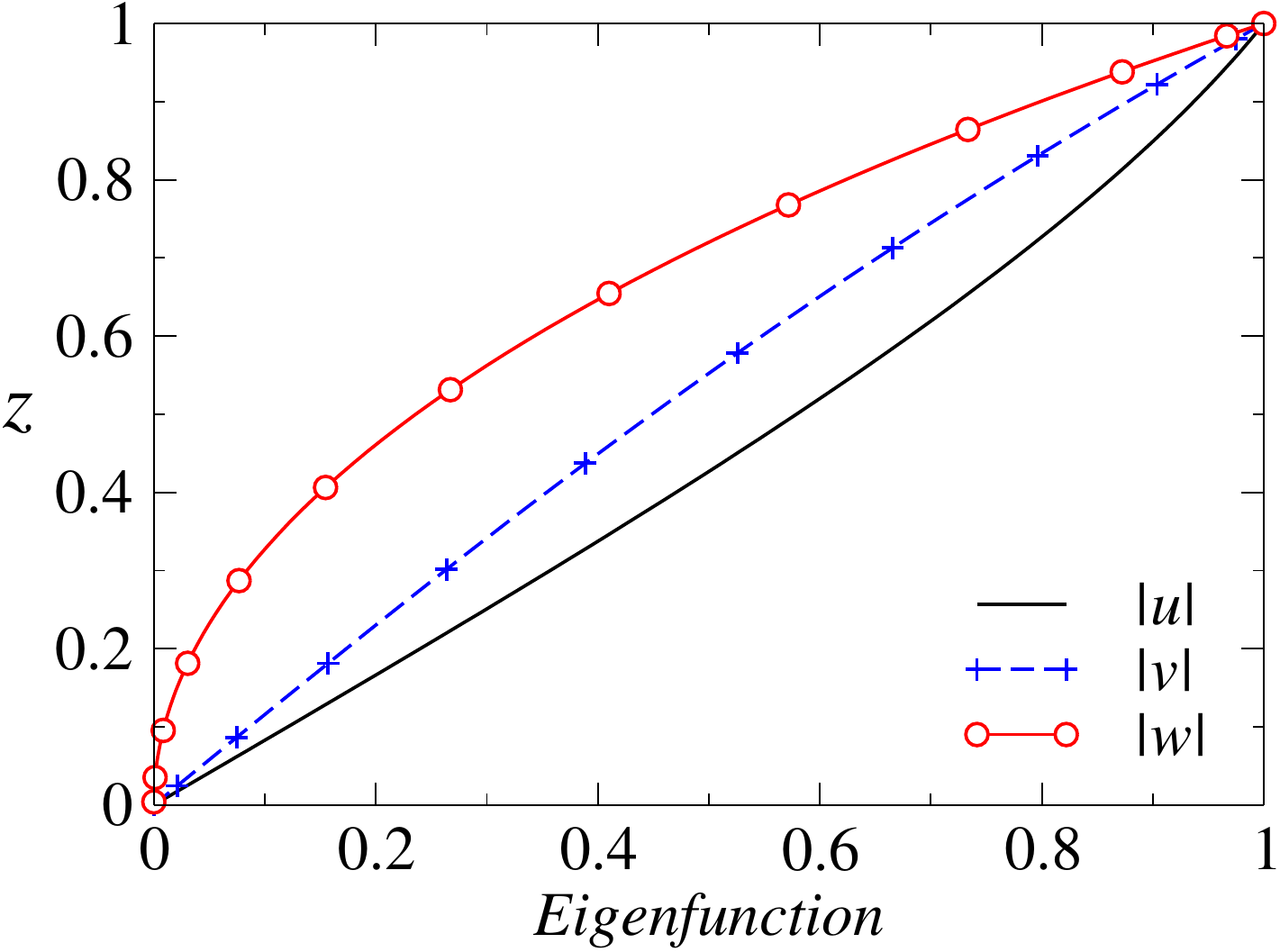} 
\caption{Variation of the absolute values of the streamwise ($|u|$), spanwise ($|v|$), and normal ($|w|$) perturbations along the $z$ direction. Here, $k = 0$, $m = 0.01$, $Re_\Omega = 0.5$, $Re = 0.1$ and $S = 0.01$.}
\label{fig:UVW}
\end{figure}

Figure~\ref{fig:DCvaryReROmvswiwr}(b) shows the variation of phase speed of the spanwise mode $\omega_r/m$ with $Re_\Omega$ for different values of $Re$. At $Re_\Omega=0$, the eigenspectrum does not contain any mode with $\omega_r=0$. As $Re_\Omega$ assumes a finite value, a new mode comes into existence possessing $\omega_r<0$ whose $\omega_i$ increases with increasing $Re_\Omega$. This implies that a whole new spanwise mode arises as a result of the combined effect of wind stress and the rotation effect. To conclude, the new streamwise mode arises due to the rotation effect destabilising an existing free-surface mode. In contrast, the new spanwise mode comes into existence due to the synergetic effect of wind stress and rotation effect. From Figures~\ref{fig:DCReOmvswiwrk001}(b) and \ref{fig:DCvaryReROmvswiwr}(b), for all values of $Re$, the curves overlap, demonstrating that $\omega_r$ is independent of $Re$ in the limit of small streamwise and spanwise wavenumbers. This numerical observation agrees with the longwave asymptotic analysis presented in Section~\ref{sec:LongWave}, which shows analytically that, for the longwave spanwise mode, the leading-order phase speed depends strongly on $Re_\Omega$ but is independent of $Re$.

\begin{figure}
\centering
\hspace{0.5mm} (a) \hspace{4.8cm} (b) \\
\includegraphics[width=0.4\textwidth]{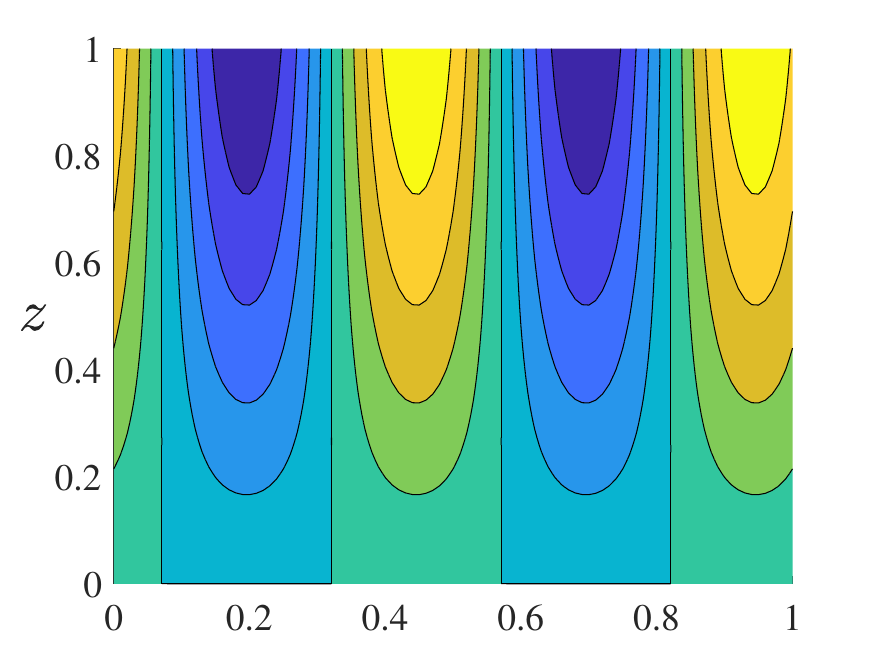} \hspace{1mm}
\includegraphics[width=0.4\textwidth]{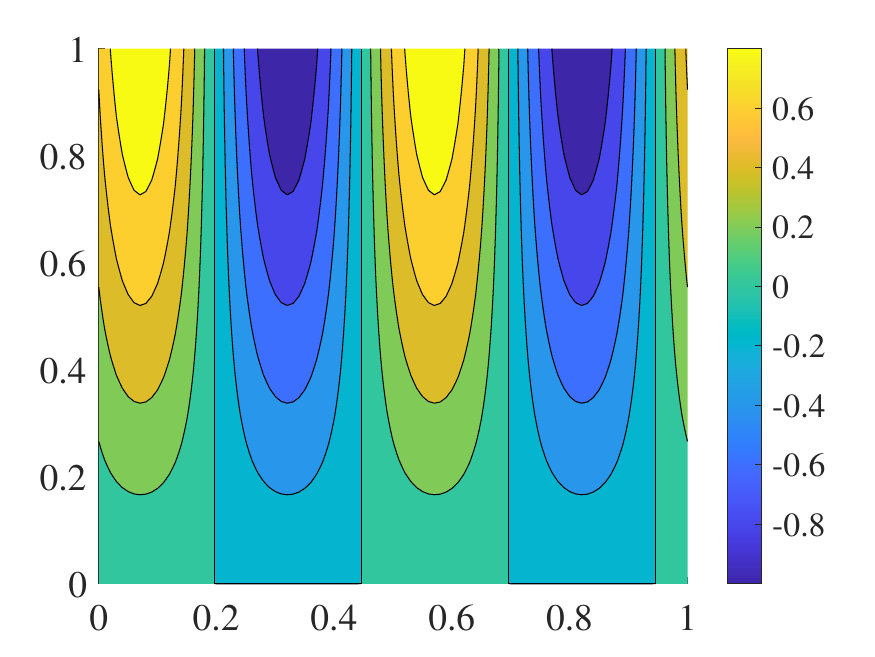}  \\
\hspace{0.5mm} (c) \hspace{4.8cm} (d) \\
\includegraphics[width=0.4\textwidth]{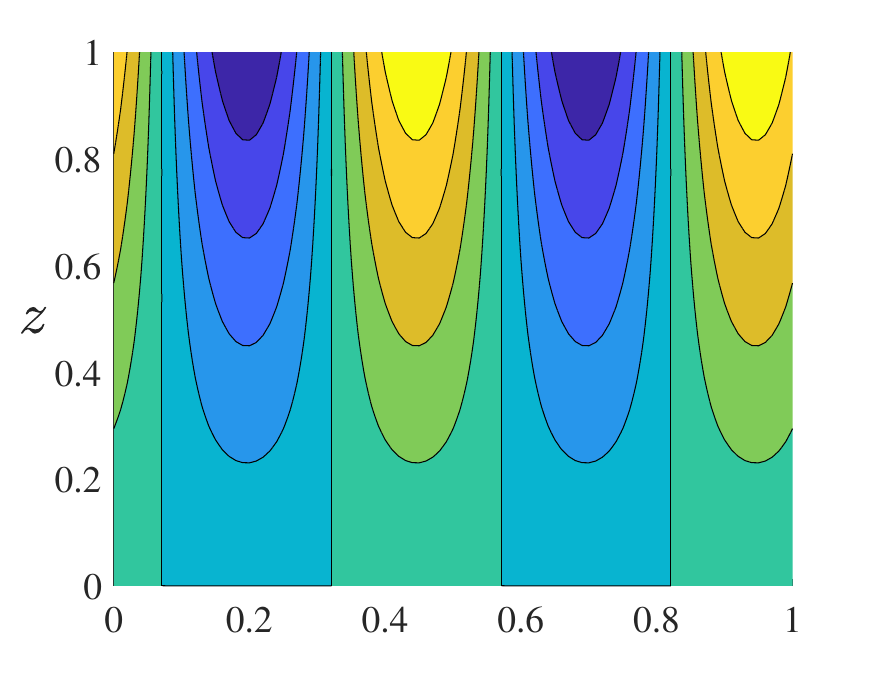} \hspace{1mm} 
\includegraphics[width=0.4\textwidth]{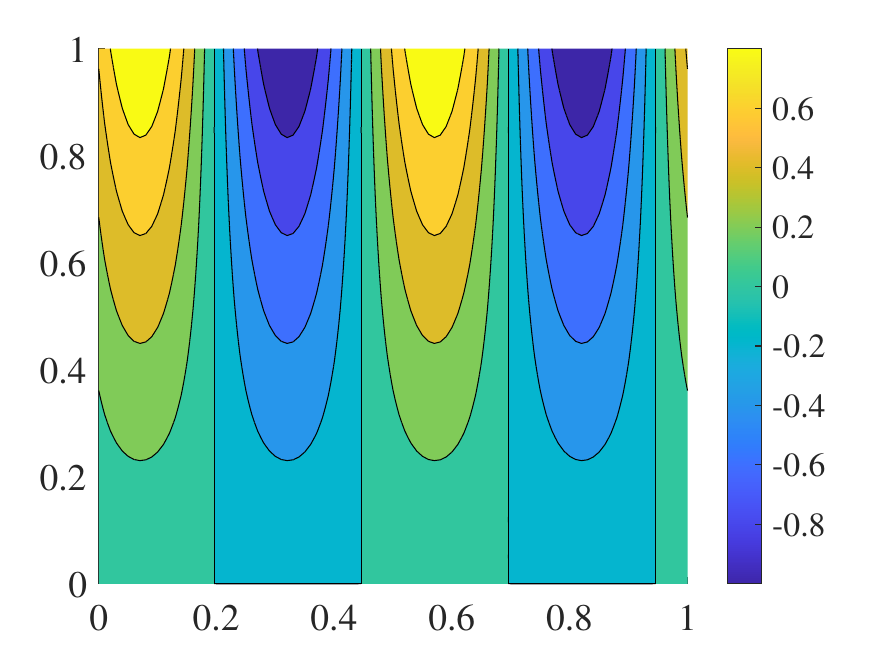}\\ 
\hspace{0.5mm} (e) \hspace{4.8cm} (f) \\
\includegraphics[width=0.4\textwidth]{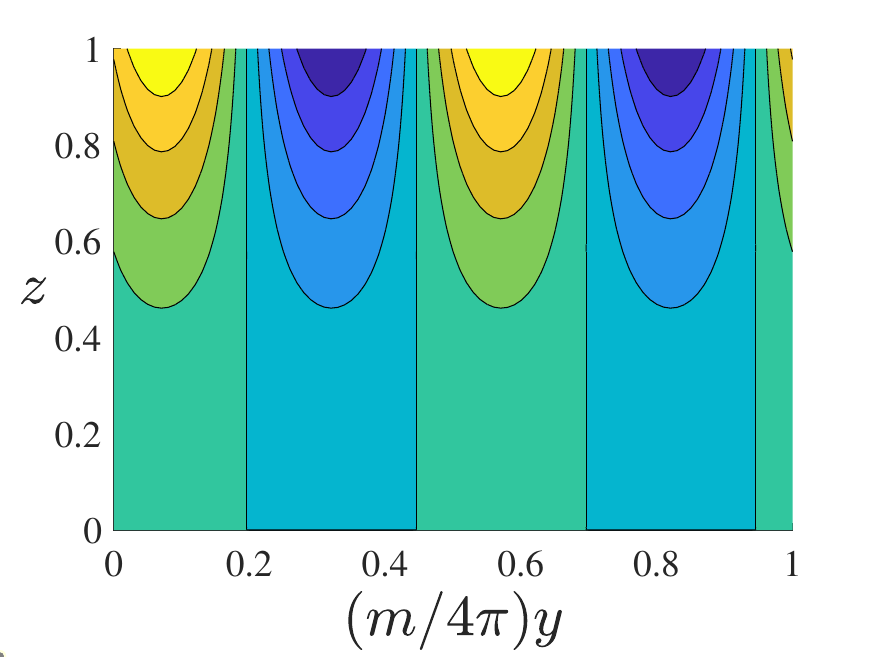} \hspace{1mm} 
\includegraphics[width=0.4\textwidth]{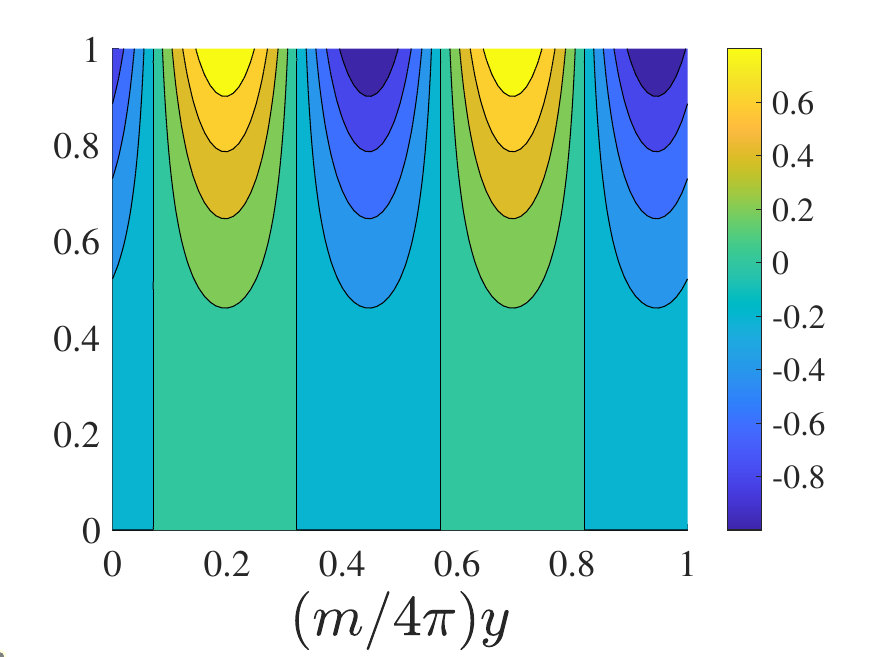}
\caption{Contours of the real $(a, c, e)$ and imaginary $(b, d, f)$ parts of the velocity perturbations ($u$, $v$, and $w$). Panels $(a, b)$, $(c, d)$, and $(e, f)$ show the contours of $u$, $v$, and $w$, respectively.  Here $k = 0$, $m=0.01$, $Re_\Omega = 0.5$, $Re = 0.1$ and $S=0.01$.}
\label{fig:Contour}
\end{figure}

To further demonstrate the destabilisation caused by rotation, we analyse the vertical variation of the eigenfunctions. Figure~\ref{fig:UVW} presents the absolute values of the perturbed velocity components in the streamwise ($u$), spanwise ($v$), and wall-normal ($w$) directions as functions of $z$ (the direction normal to the base flow). These velocity eigenfunctions correspond to the unstable mode identified in Figure~\ref{fig:DCvaryReROmvswiwr} for $k = 0$, $m=0.01$, $Re = 0.1$, $Re_\Omega = 0.5$ and $S = 0.01$. The variations of $u$, $v$, and $w$ along $z$ reveal the structure of the disturbance. It is evident that the maximum amplitude occurs at the free surface ($z = 1$), indicating that the instability originates at the free surface.  This interpretation is further supported by the contour plots of $u$, $v$, and $w$, which show significant disturbance amplitudes near the free surface, as illustrated in Figure~\ref{fig:Contour}.

\begin{figure}
\centering
\hspace{0.6cm} {\large (a)} \hspace{6.3cm} {\large (b)} \\
\includegraphics[width=0.48\textwidth]{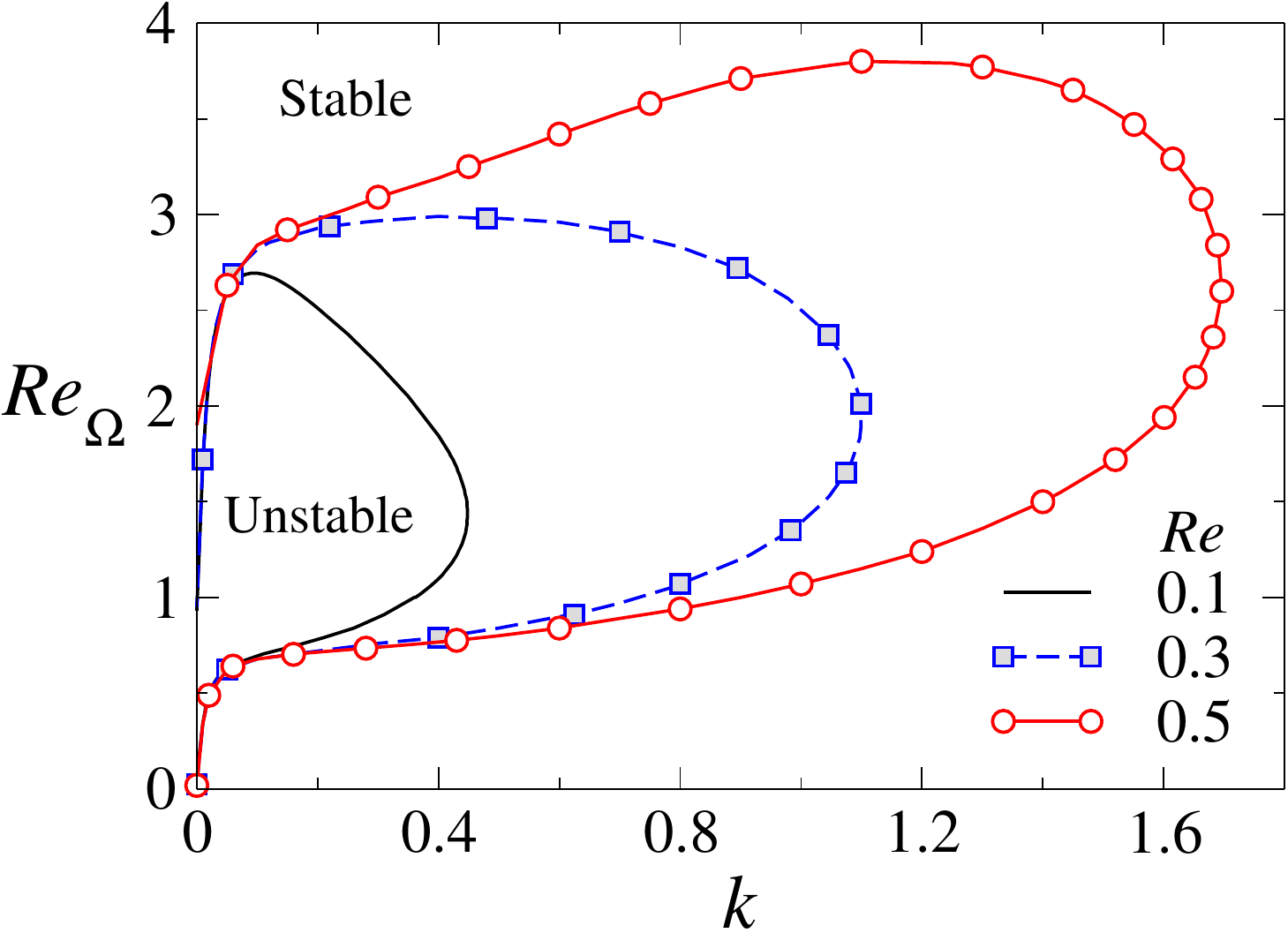} \hspace{2mm}
\includegraphics[width=0.48\textwidth]{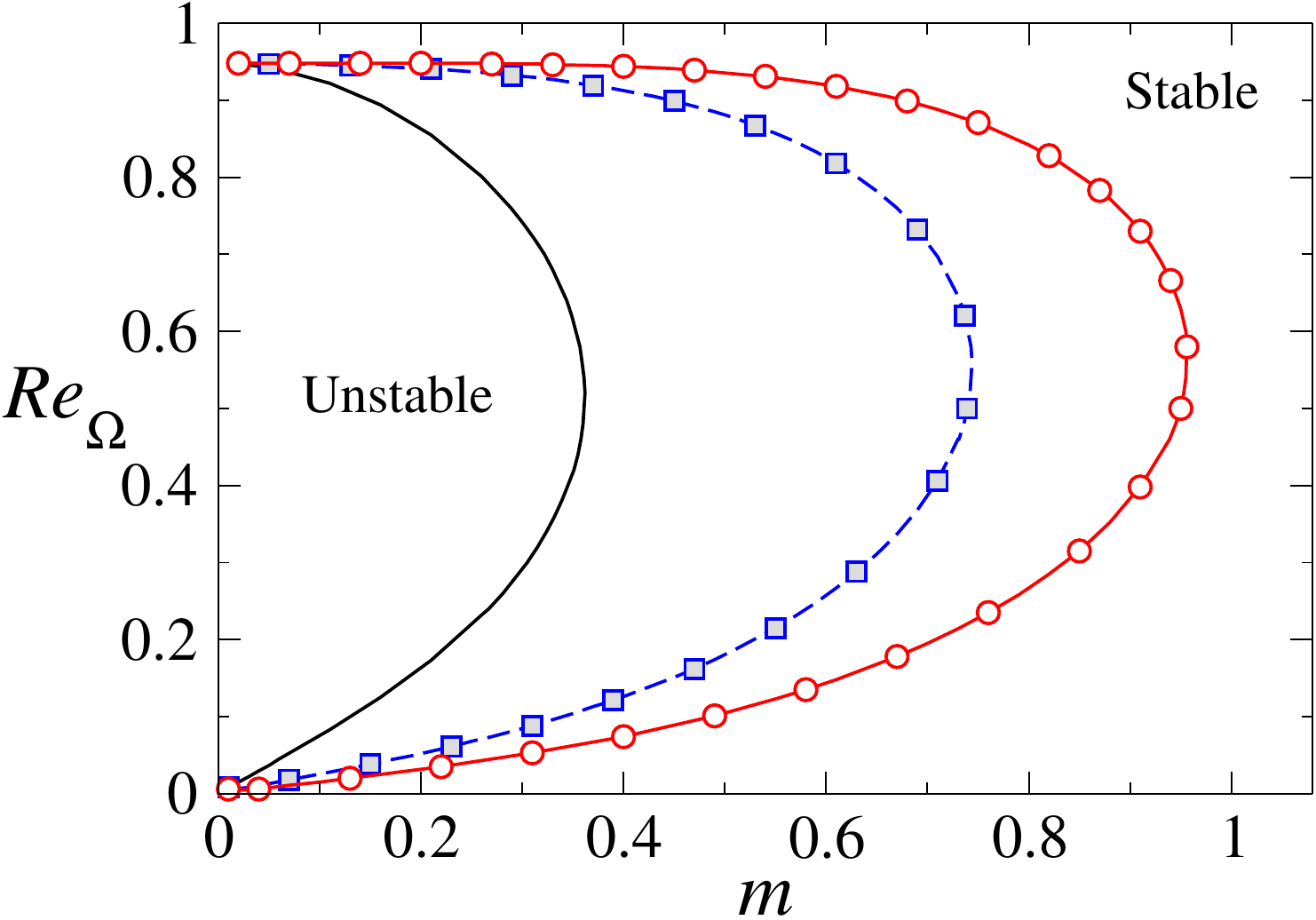}
\caption{Neutral stability curves demarcating the boundaries between stable and unstable regions: (a) in the $Re_\Omega - k$ space with $m = 0.01$, and (b) in the $Re_\Omega - m$ space with $k = 0$, for different values of $Re$. Here, $S = 0.01$.}
\label{fig:NSvaryRekmvsROm}
\end{figure}

Next, we examine the neutral stability curves (contours of $\omega_i = 0$) for various Reynolds numbers ($Re$) to understand the influence of $Re$ on stability characteristics. Figures~\ref{fig:NSvaryRekmvsROm}(a) and \ref{fig:NSvaryRekmvsROm}(b) depict the neutral curves in the $Re_\Omega - k$ and $Re_\Omega - m$ planes, respectively, for different values of $Re$. The critical value of $Re_\Omega$ at which the mode becomes unstable remains nearly the same across all $Re$, indicating that the onset of instability is largely unaffected by inertial effects. However, the detailed shape of the neutral curves exhibits a strong $Re$ dependence. This distinction is significant because, while the phase speed of the longwave mode is exactly independent of $Re$ in the asymptotic limit (Section~\ref{sec:LongWave}), the growth rate $\omega_i$ varies with $Re$, influencing the shape of the neutral stability curves. Thus, an increasing inertia exacerbates the newly predicted streamwise and spanwise instabilities.

Furthermore, Figure~\ref{fig:kmvswivaryS} in Appendix A illustrates the influence of the dimensionless surface tension $S$. The results show that increasing $S$ has a stabilising effect on the flow. However, for longwave modes (small $k$ and $m$), the low-wavenumber nature of the disturbances makes the influence of surface tension on the values of the critical parameters minimal. Computations of growth rates for both streamwise ($k$) and spanwise ($m$) disturbances indicate that the most unstable case consistently occurs for $S = 0$, as expected, since the absence of surface tension removes any restoring force opposing surface deformations, allowing disturbances to grow freely. Increasing $S$ reduces the growth rate, confirming its stabilising effect. Nevertheless, $S = 0$ is not physically realistic, as any liquid–air interface has a finite surface tension, and this limit can admit spurious shortwave modes of arbitrarily small wavelength that are suppressed when $S$ is finite. To avoid such nonphysical behaviour while remaining close to the most unstable regime, we adopt a small but finite value $S = 0.01$. This choice preserves the essential instability mechanism due to rotation and shear while ensuring physical consistency of the results.

\subsection{Longwave asymptotic analysis ($m\rightarrow0$)} \label{sec:LongWave}

The preceding discussion established the existence of new longwave streamwise and spanwise instabilities through linear stability analysis using the Chebyshev spectral collocation method. In this section, we show that the predicted spanwise instability can also be captured analytically using a longwave asymptotic approach. Classical longwave analyses \citep{yih1963, yiantsios1988linear, khomami1990interfacial, sahu2007linear}, typically consider two-dimensional perturbations by setting the spanwise velocity perturbation $v = 0$ and spanwise wavenumber $m = 0$. However, due to the inherently three-dimensional nature of the dominant mode in the present study (the spanwise mode), such restrictions cannot be applied. Therefore, we perform a fully three-dimensional perturbation expansion for all variables in terms of $m$, while setting $k = 0$. Specifically, we employ asymptotic expansions in powers of the streamwise wavenumber $m$ for the complex phase speed $c$ and the vertical velocity $w$, ensuring that the streamwise velocity $u$, spanwise velocity $v$, and pressure $p$ satisfy the continuity equation at each order of $m$. Additionally, we substitute $\omega = m c$, where $c$ represents the speed of spanwise perturbations. In the following, we present the asymptotic expansions employed for all variables in our analysis:
\begin{subequations}\label{eq:Asymptotic}
\begin{eqnarray}
u&=&\frac{1}{m}\;u_0+u_1+\cdot \cdot \cdot,\\
v&=&\frac{1}{m}\;v_0+v_1+\cdot \cdot \cdot,\\
w&=&w_0+m\; w_1+\cdot \cdot \cdot, \\
p&=&\frac{1}{m^2}\;p_0+\frac{1}{m}\;p_1+\cdot \cdot \cdot, \\
c&=&c_0+m\;c_1+\cdot \cdot \cdot. 
\end{eqnarray}
\end{subequations}

Substituting the asymptotic expansions (Eqs.~\ref{eq:Asymptotic}) into the eigenvalue problem (Eq.~\ref{eq:EigenValue}) along with the associated boundary conditions (Eq.~\ref{eq:EigenValueBCs}) produces a hierarchy of equations at successive orders of the spanwise wavenumber $m$. These equations are solved sequentially to determine the coefficients of the phase speed expansion, namely $c_0$ and $c_1$. The longwave analysis is carried out using \textsc{MATHEMATICA}$^{\circledR}$.

The leading order wave speed, $c_0$, is obtained by collecting and solving the terms at $O(1)$. This leads to the following system of equations:
\begin{subequations}{\label{eq:O(1)}}
\begin{eqnarray}
2 \;Re_\Omega \; v_0+D^2u_0=0, \\
-\i p_0-2 \; Re_\Omega \; u_0+D^2v_0=0, \\
p_0=0, \\
\i v_0+Dw_0=0.
\end{eqnarray}
\end{subequations}

\begin{figure}
\centering
\includegraphics[width=0.5\textwidth]{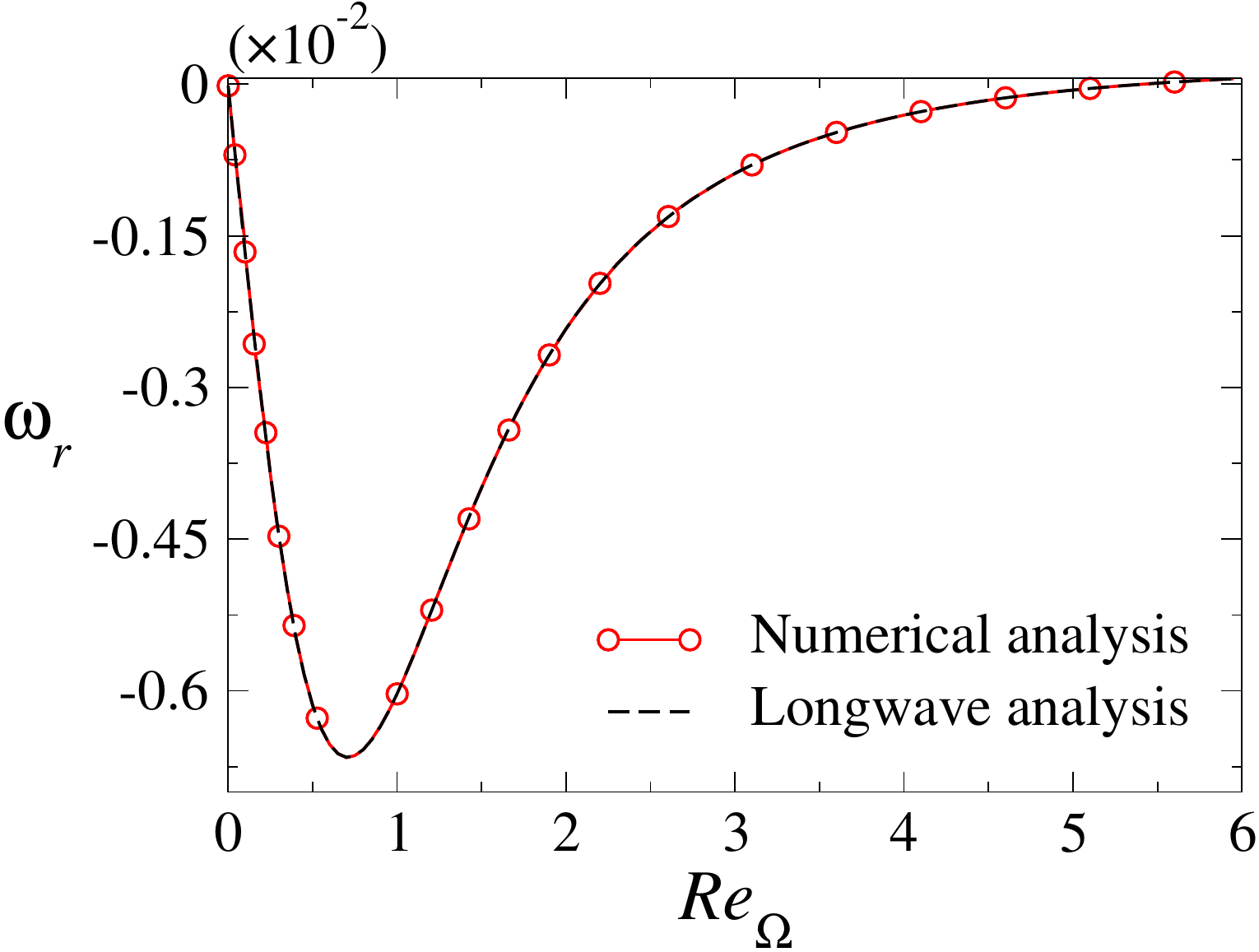} 
\caption{Comparison of the variation of the real part of the disturbance frequency $(\omega_r)$ with the rotation Reynolds number $(Re_\Omega)$ obtained using the numerical and longwave approaches. The rest of the parameters are $k = 0$, $m = 0.01$, $Re = 0.1$ and $S = 0.01$.}
\label{fig:mvsReOmLWV0}
\end{figure}

The corresponding boundary conditions for the $O(1)$ system of equations are given below. 

\begin{subequations}{\label{eq:O(1)BC}} 
~At $z=0$:
\begin{eqnarray}
u_0=v_0=w_0=0, 
\end{eqnarray}

and at $z=1$:
\begin{eqnarray}
   (c_0-\bar{v})\;p_0=0,\\
  (c_0-\bar{v})\;Du_0-2 \;\i \; Re_\Omega \; \bar{v} \; w_0=0,\\
  (c_0-\bar{v})\;Dv_0+2 \; \i \; Re_\Omega \; \bar{u} \; w_0=0.
\end{eqnarray}
\end{subequations}
These conditions enforce no-slip and no-penetration at the lower boundary ($z = 0$) and impose dynamic constraints at the free surface ($z = 1$), ensuring continuity of normal and tangential stresses in the presence of system rotation. Solving the $O(1)$ governing equations \eqref{eq:O(1)} together with the boundary conditions \eqref{eq:O(1)BC} yields the wave speed $c_0$, which is found to be real. This solution provides two expressions for $c_0$: one corresponds to the trivial solution $c_0 = \bar{v}$, and the other is given by
\begin{align}
 &c_0=\frac{1}{2 q \left[\cos(2 q) +\cosh(2 q)\right]^2}\left[\cos(3q)\sinh(q)-\cosh(3q)\sin(q)+ \nonumber  \right. \\& \left.\hspace{0.5cm} \cosh(q)\{2\sin(q)+\sin(3q)\}-\cos(q)\{2\sinh(q)+\sinh(3q)\}\right].
   \label{eq:c_0}
\end{align}
Notably the leading-order phase speed $c_0$ depends only on the rotational Reynolds number $Re_\Omega$, and is independent of the Reynolds number $Re$. This analytical result agrees with our earlier numerical findings (Figure~\ref{fig:DCvaryReROmvswiwr}b), where the instability characteristics were shown to be independent of $Re$, as the curves for different $Re$ values coincide. Therefore, in the present longwave analysis, it is sufficient to present results for a single value of $Re$. Figure~\ref{fig:mvsReOmLWV0} demonstrates that the variation of $\omega_r$ with $Re_\Omega$ obtained from the asymptotic analysis is in excellent agreement with the numerical results.

Similarly, solving the $O(m)$ equations together with the associated boundary conditions (given by Eqs.~\eqref{eq:O(k)} and \eqref{eq:O(k)BC}) yields the coefficient $c_1$, which is found to be a complex. By collecting terms at $O(m)$, we obtain the following system of equations:
\begin{subequations}{\label{eq:O(k)}}
\begin{eqnarray}
2 \;Re_\Omega \; v_1+Re \; (\i u_0 \;(c_0-\bar{v})&-&w_0 \; D\bar{u})+D^2u_1=0, \\
-\i  p_1-2 \; Re_\Omega  \; u_1+ Re \; (\i  v_0 \; (c_0-\bar{v})&-&w_0 \;D\bar{v})+D^2v_1=0, \\
p_1&=&0, \\
\i v_1&+&Dw_1=0.
\end{eqnarray}
\end{subequations}

\begin{figure}
\centering
\hspace{0.6cm} {\large (a)} \hspace{6.0cm} {\large (b)} \\
\includegraphics[width=0.49\textwidth]{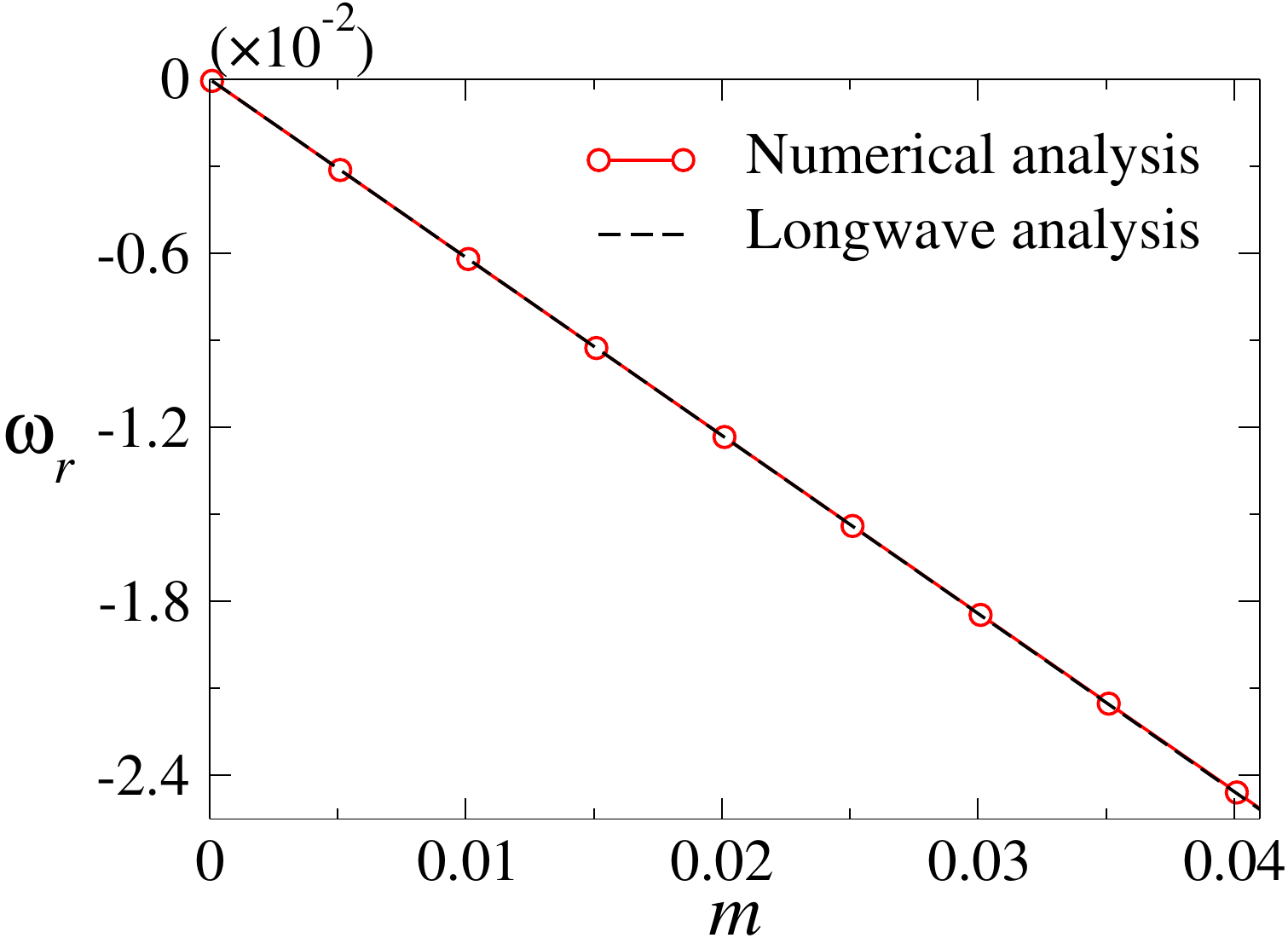}
\includegraphics[width=0.49\textwidth]{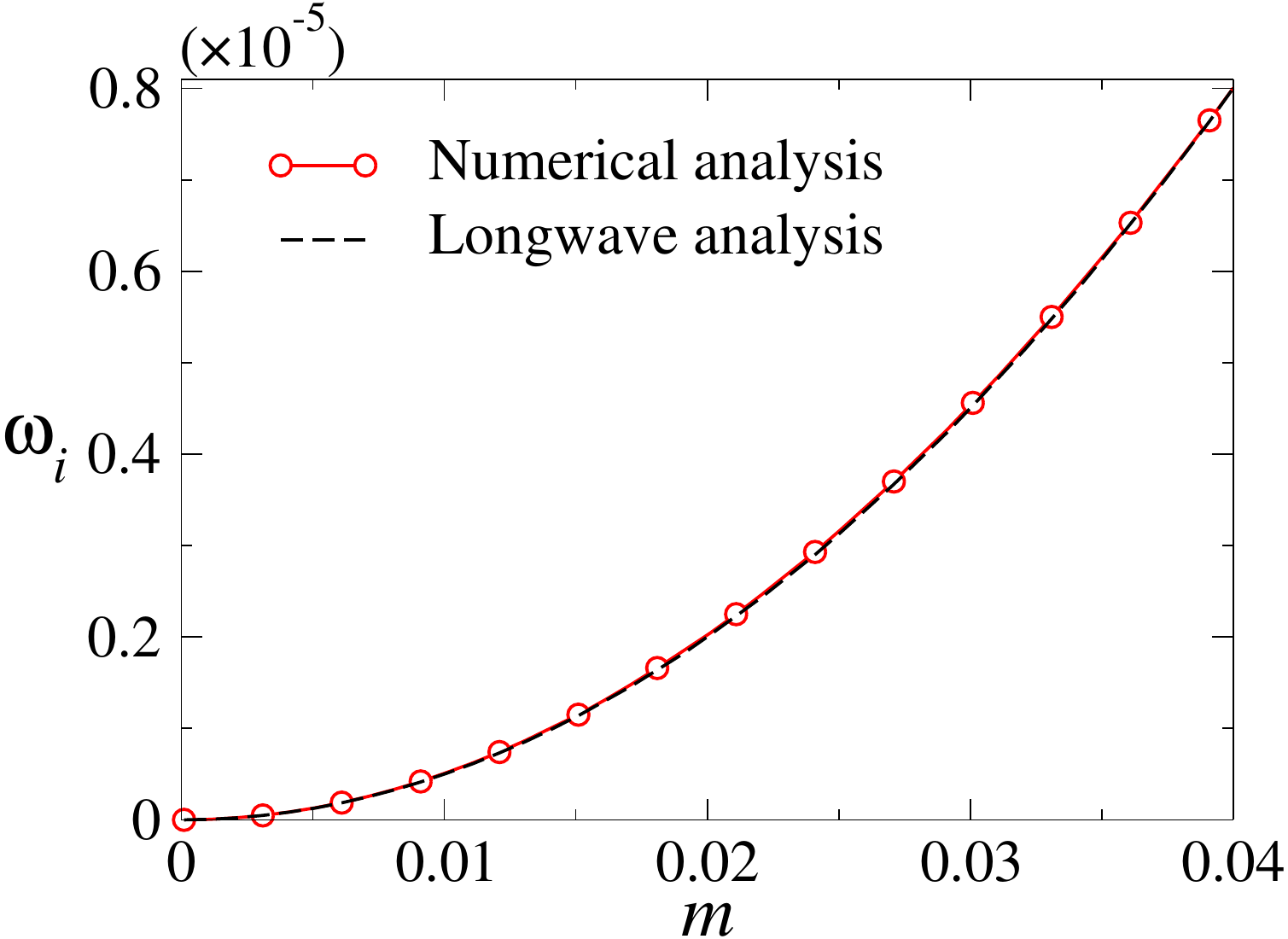}
\caption{Comparison of the variations (a) $\omega_r$ and (b) $\omega_i$ with $m$ obtained using the numerical and longwave simulations. The rest of the parameters are $k=0$, $Re=0.1$, $Re_\Omega=0.5$ and $S=0.01$.}
\label{fig:LWmvswrwrV1}
\end{figure}

The corresponding boundary conditions for $O(m)$ equations are given below. 

\begin{subequations}{\label{eq:O(k)BC}}
At $z=0$:
\begin{eqnarray}
  u_1=v_1=w_1=0, 
\end{eqnarray}
and at $z=1$:
\begin{eqnarray}
c_1 \; p_0 + (c_0 &-& \bar{v}) \; p_1=0, \\
c_1\; Du_0+(c_0-\bar{v}) \; Du_1 &-& 2 \;\i \; Re_\Omega\bar{v} \; w_1=0, \\
c_1\; Dv_0+(c_0-\bar{v})\; Dv_1 &+& 2 \; \i \;Re_\Omega\bar{u} \; w_1=0.
\end{eqnarray}
\end{subequations}
The resulting expression for $c_1$, while consistent with the numerical results, is too cumbersome to present explicitly and depends on $Re_\Omega$ and $Re$. However, for the specific values of $Re_\Omega$ corresponding to Figure~\ref{fig:LWmvswrwrV1}(b), the analytical solution for the phase speed $c_1$ is found to be
\begin{eqnarray}
c_1=(0.0507691 \; \i)\;Re.
\end{eqnarray}
The Mathematica code used to obtain the expression for $c_1$ for a specific set of parameters is provided in the supplementary information. In the numerical formulation, we reported the complex frequency $\omega=\omega_r+\i\omega_i$. For the spanwise longwave expansion with $k=0$, we have 
\begin{eqnarray}
     \omega=mc=m(c_0+mc_1),
\end{eqnarray}
where $c_0$ is purely real and $c_1$ is complex. It then follows that
\begin{eqnarray}
     \omega_r=mc_0+m^2\;\text{Real}(c_1), \quad \text{and}  \quad \omega_i=m^2\;\text{Imag}(c_1).
\end{eqnarray}
To validate the asymptotic results, we compare them with numerical solutions obtained using the spectral collocation method. Figures~\ref{fig:mvsReOmLWV0} and \ref{fig:LWmvswrwrV1} present this comparison, showing the numerically computed eigenvalues alongside predictions from the longwave asymptotic theory. In Figures~\ref{fig:LWmvswrwrV1}(a) and \ref{fig:LWmvswrwrV1}(b), the real and imaginary parts of the perturbation frequency, $\omega_r$ and $\omega_i$, respectively, are plotted as functions of the spanwise wavenumber $m$. The dashed black lines denote the asymptotic predictions, obtained by plotting $\omega_r$ and $\omega_i$ against $m$. The asymptotic results show excellent agreement with the numerical solutions in the longwave limit, confirming that the predicted new longwave spanwise instability is genuine and not a numerical artefact. Thus, the asymptotic analysis serves two purposes. First, in the absence of prior studies on the combined effects of shear and rotation, it validates our linear stability solver through comparison with the asymptotic results. Second, it confirms the existence of the newly identified longwave instability reported in this study.
 
\section{Concluding remarks} \label{sec:conc}

We have performed a linear stability analysis of a shear flow driven by wind stress in the presence of rotational effects, relevant to ocean wave dynamics. A constant wind shear stress of magnitude $\tau$ is applied at the free surface, and the lower plane is rotating at a steady angular velocity $\Omega$. The frame of reference is fixed to the Earth. The rotational influence is quantified by the Taylor number ($\Ta$), which appears in the definition of the rotational Reynolds number ($\Re_\Omega = \Re \cdot \Ta$), with $Re$ denoting the Reynolds number. We investigate the influence of the rotational Reynolds number ($\Re_\Omega$) and the Reynolds number ($Re$) on the stability characteristics of the flow. The resulting eigenvalue problem is solved using the Chebyshev spectral collocation method. To validate our numerical approach, we reproduce the neutral stability curve for the non-rotating case, which agrees well with the results of \citet{smith1982instability}.

The linear stability analysis reveals new longwave instability modes. Depending on the wavenumber, these unstable modes are classified as streamwise ($m=0$) and spanwise ($k=0$) modes. Our results show that the spanwise longwave mode is the most unstable, exhibiting the lowest $Re_\Omega$ at the onset of instability. In fact, the critical value is $Re_\Omega=0$ for the spanwise mode, implying that the flow becomes unstable for $Re_\Omega > 0$. The eigenfunctions and velocity perturbation contours demonstrate that rotation destabilises the streamwise free-surface mode. The dominant spanwise mode originates from the combined influence of rotation and the shear imposed at the free surface. Owing to the longwave nature of the spanwise mode, we employ a longwave asymptotic analysis to capture the newly identified instability for $k=0$ analytically. Unlike conventional longwave analyses, which typically restrict attention to two-dimensional perturbations, the present problem necessitates the inclusion of three-dimensional disturbances, resulting in significantly more intricate mathematical formulations. Despite this complexity, our asymptotic analysis accurately predicts the most unstable eigenvalue and exhibits excellent agreement with numerical results at small values of the spanwise wavenumber of the disturbance, $m$. This ascertains that the predicted instability is genuinely present and not a numerical artefact. In the context of ocean wave dynamics, this suggests that the longwave instability could be induced by the combined action of Earth's rotation and shear stress exerted by the wind, which can potentially lead to highly unstable ocean waves. 

\appendix 
\label{appendix}

\section{Effect of surface tension}

\begin{figure}
\centering
\hspace{0.6cm} {\large (a)} \hspace{6.0cm} {\large (b)} \\
\includegraphics[width=0.45\textwidth]{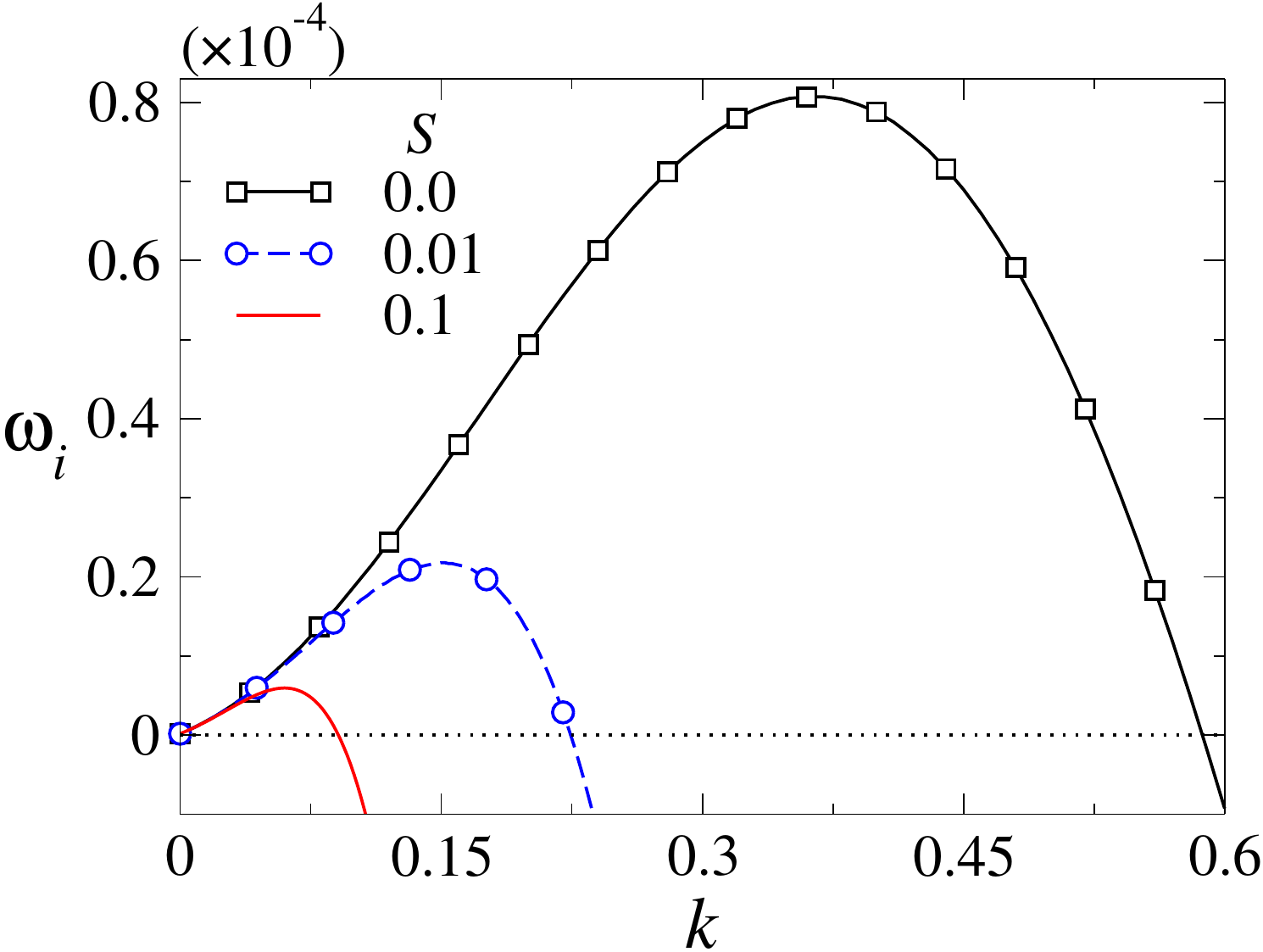} \hspace{2mm}
\includegraphics[width=0.45\textwidth]{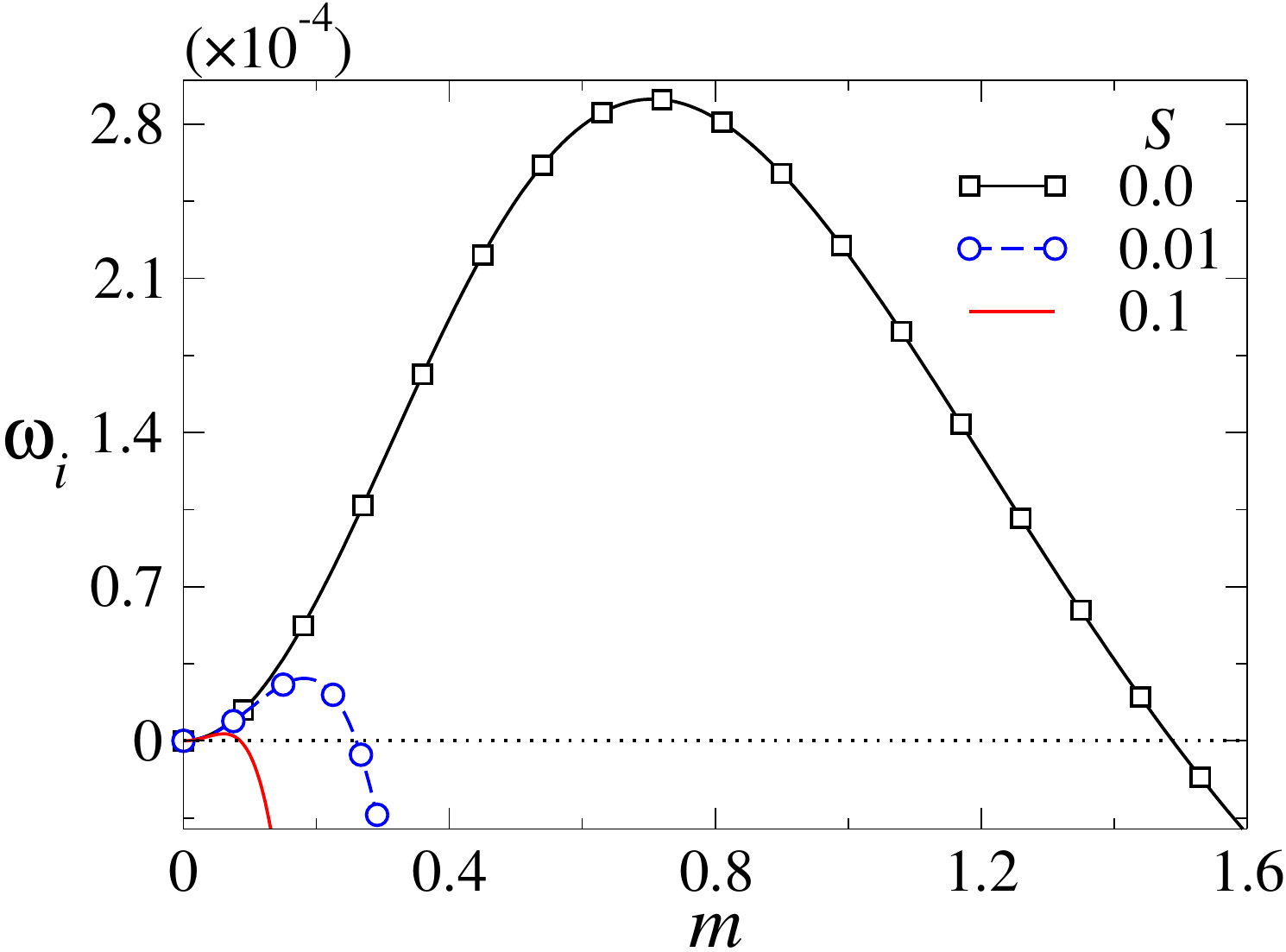}
\caption{Variations of (a) $\omega_i$ versus $k$ with $m=0.01$ and (b) $\omega_i$ versus $m$ with $k=0$ for different values of $S$. The rest of the parameters are $Re_\Omega=0.8$ and $Re = 0.1$.}
\label{fig:kmvswivaryS}
\end{figure}

Figures~\ref{fig:kmvswivaryS}(a) and \ref{fig:kmvswivaryS}(b) examine the effect of the surface tension parameter $S$ on the dispersion curves, showing $\omega_i$ versus $k$ for $m = 0.01$ and $\omega_i$ versus $m$ for $k = 0$, respectively. These curves are obtained from our linear stability analysis. The results indicate that the instability is strongest for $S = 0$ and that increasing $S$ progressively stabilizes the flow, consistent with earlier observations by \cite{smith1982instability}. For very small wavenumbers ($k \approx 0$ and $m \approx 0$), the influence of surface tension is negligible, and the dominant modes remain largely unaffected. Although $S = 0$ corresponds to the most unstable case, this limit is not physically realistic, as every free surface possesses a finite surface tension. To remain close to the most unstable regime while maintaining physical consistency, we adopt a small but finite value $S = 0.01$ throughout this study. This choice allows a clear understanding of the underlying instability mechanisms influenced by rotation. 

\clearpage

\noindent{\bf Declaration of Interests:} The authors report no conflict of interest. \\
\\
\noindent{\bf Acknowledgment:} {K. C. S. thanks IIT Hyderabad for the financial support through the grant IITH/CHE/F011/SOCH1. R.P. acknowledges financial support from the SERB-DST under grant SRG/2023/000223. We are deeply grateful to the anonymous reviewers and to Prof. James T. Kirby for providing valuable suggestions that enhanced the quality and depth of the manuscript.}


\end{document}